\documentclass[letterpaper,twocolumn,10pt]{article}
\usepackage{usenix}
\usepackage{enumerate}
\usepackage{amsmath}
\usepackage{amssymb}
\usepackage{tabularx}
\usepackage{pgfplots}
\usepackage{pdfpages}
\usepackage{rotating}
\usepackage{hyperref}

\usepackage{listings}
\usepackage{xcolor}
\usepackage[toc, acronyms]{glossaries}
\usepackage{cleveref}
\usepackage[justification=centering]{caption}
\usepackage{tikz}
\usepackage{todonotes}
\usepackage{xspace}
\usepackage{enumitem}

\usepackage[english]{babel}
\addto\extrasenglish{

}

\newcommand*{\field}[1]{\mathbb{#1}}%

\definecolor{codegreen}{rgb}{0,0.6,0}
\definecolor{codegray}{rgb}{0.5,0.5,0.5}
\definecolor{codepurple}{rgb}{0.58,0,0.82}
\definecolor{backcolour}{rgb}{0.95,0.95,0.92}
\lstdefinestyle{sourcecode}{
    backgroundcolor=\color{backcolour},   
    commentstyle=\color{codegreen},
    keywordstyle=\color{magenta},
    numberstyle=\tiny\color{codegray},
    stringstyle=\color{codepurple},
    basicstyle=\ttfamily\footnotesize,
    breakatwhitespace=false,         
    breaklines=true,                 
    captionpos=b,                    
    keepspaces=true,                 
    numbers=left,                    
    numbersep=5pt,                  
    showspaces=false,                
    showstringspaces=false,
    showtabs=false,                  
    tabsize=2
}
\newacronym{sca}{SCA}{Side Channel Attack}
\newacronym{vfi}{VFI}{Voltage Fault Injection}
\newacronym{mvfi}{MVFI}{Multiple Voltage Fault Injection}
\newacronym{mlfi}{MLFI}{Multiple Laser Fault Injection}
\newacronym{mcfi}{MCFI}{Multiple Clock Fault Injection}
\newacronym{fi}{FI}{Fault Injection}
\newacronym{dut}{DuT}{Device under Test}
\newacronym{ic}{IC}{Integrated Circuit}
\newacronym{tee}{TEE}{Trusted Execution Environment}
\newacronym{ree}{REE}{Rich Execution Environment}
\newacronym{mpu}{MPU}{Memory Protection Unit}
\newacronym{mfi}{MFI}{Multiple Fault Injection}
\newacronym{sfi}{SFI}{Single Fault Injection}
\newacronym{tzm}{TZM}{TrustZone-M}
\newacronym{pcb}{PCB}{Printed Circuit Board}
\newacronym{sdk}{SDK}{Software Development Kit}
\newacronym{fpga}{FPGA}{Field Programmable Gate Array}
\newacronym{emfi}{EMFI}{Electromagnetic Fault Injection}
\newacronym{mcu}{MCU}{Microcontroller Unit}
\newacronym{ecc}{ECC}{Error Correcting Codes}
\newacronym{drm}{DRM}{Digital Rights Management}
\newacronym{fm}{FM}{Fault Model}
\newacronym{poc}{PoC}{Proof-of-Concept}
\newacronym{sau}{SAU}{Security Attribution Unit}
\newacronym{idau}{IDAU}{Implementation Defined Attribution Unit}
\newacronym{lsb}{LSB}{Least Significant Bit}
\newacronym{api}{API}{Application Programming Interface}
\newacronym{usb}{USB}{Universal Serial Bus}
\newacronym{hdl}{HDL}{Hardware Description Language}
\newacronym{ahb}{AHB}{Advanced High-Performance Bus}
\newacronym{sf}{SF}{Success Function}
\newacronym{psf}{PSF}{Partial Success Function}
\newacronym{des}{DES}{Data Encryption Standard}
\newacronym{aes}{AES}{Advanced Encryption Standard}
\newacronym{dpa}{DPA}{Differential Power Analysis}
\newacronym{cpa}{CPA}{Correlation Power Analysis}
\newacronym{rsa}{RSA}{Rivest-Shamir-Adleman-Cryptosystem}
\newacronym{ilc}{ILC}{Information Level Countermeasures}
\newacronym{hlc}{HLC}{Hardware Level Countermeasures}
\newacronym{lfi}{LFI}{Laser Fault Injection}
\newacronym{pd}{PD}{Phase Detector}
\newacronym{pll}{PLL}{Phase Locked Loop}
\newacronym{iic}{I\textsuperscript{2}C}{Inter Integrated Circuit}
\newacronym{dfa}{DFA}{Differential Fault Analysis}
\newacronym{soc}{SoC}{System-on-Chip}
\newacronym{os}{OS}{Operating System}
\newacronym{gcc}{GCC}{Gnu Compiler Collection}
\newacronym{ide}{IDE}{Integrated Development Environment}
\newacronym{gpio}{GPIO}{General Purpose Input/Output}
\newacronym{ldo}{LDO}{Low Dropout Regulator}
\newacronym{sfu}{SFU}{Single Fault Unit}
\newacronym{sgx}{SGX}{Software Guard Extension}
\newacronym{dac}{DAC}{Digital to Analog Converter}
\newacronym{bod}{BoD}{Brownout Detection}
\newacronym{bo}{BO}{Buffer Overflow}
\newacronym{ace}{ACE}{Arbitrary Code Execution}
\newacronym{dma}{DMA}{Direct Memory Access}
\newacronym{gnd}{GND}{Ground}
\newacronym{gtzc}{GTZC}{Global TrustZone Controller}
\newacronym{por}{POR}{Power-On Reset}
\newacronym{apb}{APB}{Advanced Peripheral Bridge}
\newacronym{mc}{MC}{Memory Checker}
\newacronym{pc}{PC}{Peripheral Checker}
\newacronym{wu}{WU}{Wrapper Unit}
\newacronym{puf}{PUF}{Physical Unclonable Function}
\newacronym{trng}{TRNG}{True Random Number Generator}
\newacronym{drbg}{DRNG}{Deterministic Random Bit Generator}
\newacronym{tz}{TZ}{TrustZone}
\newacronym{dvfs}{DVFS}{Dynamic Voltage and Frequency Scaling}
\newacronym{rop}{ROP}{Return Oriented Programming}
\newacronym{cfi}{CFI}{Clock Fault Injection}
\newacronym{ram}{RAM}{Random Access Memory}
\newacronym{rom}{ROM}{Read-only Memory}
\newacronym{llvm}{LLVM}{Low Level Virtual Machine}
\newacronym{mosfet}{MOSFET}{metal–oxide–semiconductor field-effect transistor}
\newacronym{sev}{SEV}{Secure Encrypted Virtualization}
\newacronym{sp}{SP}{Secure Processor}
\newacronym{pac}{PAC}{Peripheral Access Control}
\newcommand*{\coolname}{$\mu$-Glitch\xspace}

\begin{document}

\date{}
\title{\Large \bf Oops..! I Glitched It Again!\\
How to Multi-Glitch the Glitching-Protections on ARM TrustZone-M}
\author{
{\rm Xhani Marvin Saß}\\
Technical University of Darmstadt
\and
{\rm Richard Mitev}\\
Technical University of Darmstadt
\and
{\rm Ahmad-Reza Sadeghi}\\
Technical University of Darmstadt
}
\maketitle

\begin{abstract}
\gls*{vfi}, also known as power glitching, has proven to be a severe threat to real-world systems. 
In \gls*{vfi} attacks, the adversary disturbs the power-supply of the target-device forcing the device to illegitimate behavior.  
Various countermeasures have been proposed to address different types of fault injection attacks at different abstraction layers, either requiring to modify the underlying hardware or software/firmware at the machine instruction level. Moreover, only recently, individual chip manufacturers have started to respond to this threat by integrating countermeasures in their products. Generally, these countermeasures aim at protecting against \emph{single} fault injection (SFI) attacks, since \gls*{mfi} is believed to be challenging and sometimes even impractical.

In this paper, we present \coolname, the first Voltage Fault Injection (VFI) platform which is capable of injecting \emph{multiple}, coordinated voltage faults into a target device, requiring only a single trigger signal. We provide a novel flow for \gls*{mvfi} attacks to significantly reduce the search complexity for fault parameters, as the search space increases exponentially with each additional fault injection.
We evaluate and showcase the effectiveness and practicality of our attack platform on four real-world chips, featuring TrustZone-M: 
The first two have interdependent backchecking mechanisms, while the second two have additionally integrated countermeasures against fault injection. Our evaluation revealed that \coolname can successfully inject four consecutive faults within an average time of one day.
Finally, we discuss potential countermeasures to mitigate VFI attacks and additionally propose two novel attack scenarios for \gls*{mvfi}.
\end{abstract}

\section{Introduction}\glsresetall\label{sec:intro}

\gls*{fi} has proven to form a powerful threat to various computing platforms.
All fault injection methods temporarily disturb the physical runtime environment of the \gls*{dut} to cause specific misbehavior. 
Common \gls*{fi} attacks are, e.g., conducted by disturbing the supply voltage~\cite{5412860}, generating malicious clock signals~\cite{10.1145/3338508.3359577}, rapidly changing the electromagnetic environment~\cite{7004182}, or inducing a light pulse at the decapsulated \glspl*{ic}~\cite{6076471}.
The possible consequences from injecting a certain type of fault are described by a \gls*{fi} method's Fault Model.
Depending on the specific \gls*{fi} method used, the corresponding Fault Model may, e.g., be defined as 
skipping of machine instructions~\cite{238594}, 
corrupting the instruction decoding~\cite{7774479} 
or altering the data stored on a device's internal memory~\cite{6623558}.
Hence, \gls*{fi} attacks are capable of introducing vulnerabilities.
As an example, this kind of attacks have been successfully launched on \glspl*{tee}~\cite{206182, Nashimoto_Suzuki_Ueno_Homma_2021}, 
embedded devices~\cite{o2016fault, 8167705}, smart cards~\cite{6076471} 
and recently even against workstation processors~\cite{263816, 251590}.

A popular class of \gls*{fi} and the focus of this paper is \gls*{vfi}, as this approach is very versatile while exhibiting a high impact attack vector. \gls*{vfi} disturbs a \gls*{dut}'s power supply to provoke a specific malfunction.
To address \gls*{vfi} attacks, a variety of software-~\cite{10.1145/1873548.1873555, Sakamoto2021, 2014, 10.1145/2858930.2858931} as well as hardware-based~\cite{lyg2359, 23428376545267834} countermeasures have been proposed over the years.
Fortunately, the industry has recognized the severity of fault injection attacks and individual manufacturers are reacting to it by integrating countermeasures into their products.
For instance, NXP recently released multiple ARMv8-M series \glspl*{mcu}, which feature an instruction-level \gls*{fi} countermeasure to protect some of their security-critical registers from \gls*{fi} attacks.
Moreover, ARMv8-M \glspl*{mcu} that feature TrustZone-M (e.g., STM, Atmel, NXP) are equipped with a novel hardware unit on the internal bus system, performing additional checks on every bus access, in order to ensure the security properties of the \gls*{tee}. Although not explicitly aimed to mitigate \gls*{fi} attacks, this backchecking mechanism has complex interdependencies which make \gls*{fi} attacks targeting TrustZone-M, highly difficult~\cite{123698251376}.
Typically, \gls*{fi} attacks and countermeasures are dedicated to \gls*{sfi}, although \gls*{mfi} seems to be much more powerful. In \gls*{mfi}, multiple, coordinated faults of a certain type are injected after a single synchronizing \texttt{trigger} signal, in order to attack multiple target instructions during a single execution. \gls*{mfi} could theoretically be used against instruction-level based countermeasures~\cite{2014, 10.1145/1873548.1873555}, however, as stated by previous work conducting those attacks, especially \gls*{mvfi} are considered highly impractical due to the lack of precise and affordable \gls*{mfi} tools~\cite{10.1145/1873548.1873555, 123698251376} and efficient parameter search algorithms. Even though, commercial equipment for \gls*{mfi} is available, devices from e.g., Alphanov\footnote{\url{https://www.alphanov.com/en/products-services/double-laser-fault-injection}} and Riscure\footnote{\url{https://www.riscure.com/blog/security-highlight-multi-fault-attacks-are-practical}} were shown only to conduct \gls*{mlfi} which is much more resource intensive than \gls*{mvfi}.
In addition, off-the-shelf \gls*{vfi} devices from NewAE~\cite{o2014chipwhisperer} are not capable of injecting multiple faults based on a single trigger.
In this paper we address this open problem by providing a highly precise \gls*{mvfi} tool and the corresponding efficient parameter search algorithms, which enable an adversary to inject multiple, coordinated and consecutive voltage faults into any target device in order to attack 
any software-based \gls*{sfi} protection on the instruction-level.
We show that our tool is able to successfully perform a parameter search for up to four consecutive voltage faults and perform \gls*{mvfi} to skip the corresponding instructions in about one day.

To realize this, we had to overcome a number of challenges:
First, to the best of our knowledge, there exists no tool to conduct \gls*{mvfi}, so we designed and built our framework, coined \coolname.
Second, the timely effort required to search for multiple fault parameters grows exponentially with each additional fault, if a traditional fault parameter search (i.e., an exhaustive search) is used.
Hence, we designed, implemented and evaluated novel approaches, to search for multiple fault injection parameters, which are efficient enough for \gls*{mvfi} setups.

Even though our approach can be used to attack arbitrary devices, we focus on attacking \gls*{tzm}, as \glspl*{tee} form highly secure targets, which when compromised have shown to lead to the disclosure of highly sensitive information~\cite{10.1145/3319535.3354197}.
While principally \coolname can defeat many \gls*{vfi} research proposal countermeasures, as discussed in \autoref{sec:related}. However, we evaluate \coolname on four real-world example \glspl*{mcu} as most academic proposals are not open sourced. As a \gls*{poc} we attack two \gls*{tzm} \glspl*{mcu} which have protections explicitly protecting against \gls*{fi} attacks. Therefore, it cannot be successfully attacked by means of \gls*{sfi}. We also attack two other \glspl*{ic} with a subset of protections.
Our main contributions are as follows:
\begin{description}[leftmargin=0cm]

    \item[Novel \gls*{mfi} Framework] We present \coolname, a novel fault injection framework, which is capable of injecting multiple, coordinated voltage faults into any \gls*{dut}, in order to overcome \gls*{fi} countermeasures implemented on the instruction level.
    
    \item[Parameter Search Algorithm]
    We explore the impact of additional faults on the search space spanned by the combinations of multiple fault's parameters and present a novel and effective multiple fault parameter search to overcome the exponential increase in needed resources introduced by conventional algorithms.
    Our approach exhibits a 50 times speedup when searching for two consecutive voltage fault's parameters, and an 8.3 time speedup when searching for four voltage fault's parameters.
    
    \item[Real-world Attack]
    We use \coolname in order to inject multiple voltage faults into NXP's \texttt{LPC55SXX} and \texttt{RT6XX} \gls*{mcu} and hereby successfully circumventing all \gls*{fi} protections.
    By this, we are able to fully compromise the security introduced by \gls*{tzm}, by accessing the secure memory from within non-secure firmware.
    We show that this attack can be performed within an average time of one day. Further, we show that other \gls*{tzm} \glspl*{ic} can be broken using a subset of faults required for NXP \glspl*{ic}.
    
    \item[Possible Countermeasures]
    We propose a software level enhancement to the existing countermeasures, which is capable to protect from \gls*{mfi} attacks, by eliminating the possibility to search for \gls*{mfi} parameters.
    
\end{description}
\paragraph{Responsible Disclosure}
The results of this work have been responsibly disclosed to NXP Semiconductors Ltd. A response acknowledging our findings has been received.
In follow up communication the authors are collaborating with NXP Semiconductors Ltd. on finding a security patch.
The authors would like to thank NXP Semiconductors Ltd. for their timely and professional communication following the responsible disclosure of our findings.

\section{Background}\glsresetall
In the following we provide the background necessary in order to understand this work.

\subsection{Voltage Fault Injection} 
\gls*{vfi} is a specific \gls*{fi} method to inject disturbances into the power supply line of an \gls*{ic} and hence, violates the \gls*{ic}'s specified operating conditions for a certain, controlled period of time.
Most \glspl*{ic}, like \glspl*{mcu}, expect their supply voltage to be stable and steady, i.e., there should be no point during runtime, at which the supply voltage is interrupted or leaves a specified operating range.

\label{sec:voltage_fault}
\begin{figure}[h!]
    \centering
    \includegraphics[width=0.45\textwidth]{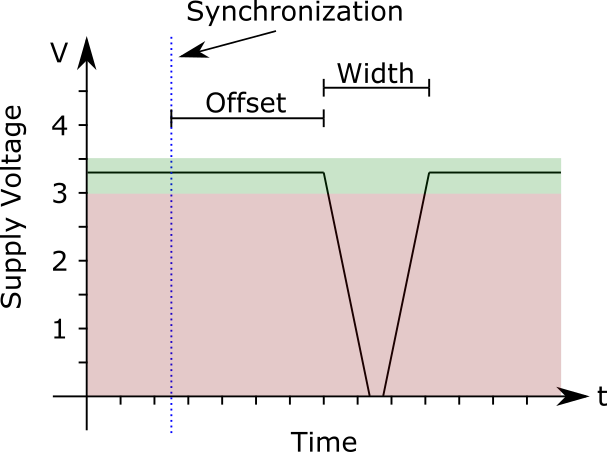}
    \caption{Example of a Voltage Fault}
    \label{fig:voltage_fault}
\end{figure}

\autoref{fig:voltage_fault} depicts a single voltage fault injected into the supply line of an \gls*{ic} running at $3.3V$.
Here the specified operating supply voltage range is highlighted green ($3.0V$ to $3.5V$), whereas the red highlighted range ($0.0V$ to $2.9V$) indicates, that the operating conditions are violated.
The fault voltage level is most commonly defined as the \gls*{gnd} reference, however, this may be optimized with respect to either reliability and repeatability~\cite{Bozzato_Focardi_Palmarini_2019}
or timely resolution~\cite{kudera2018design}.
A voltage fault is parameterized by it's \texttt{Offset} w.r.t. a synchronization point, and it's \texttt{Width}~\cite{o2016fault}.
In every \gls*{vfi} experiment the most complex part is to find the best fault parameters, in order to provide a \emph{reliable} and \emph{repeatable} attack~\cite{10.1007/978-3-319-08302-5_16}.

\vspace{-13pt}
\paragraph{The Fault Model} of \gls*{vfi} on \glspl*{mcu} consists of four different behaviors, which all arise from the effects of \gls*{vfi} on the processors internal pipeline stages namely, skipping of machine instructions, corrupting data fetches, corrupting instruction decodes and corrupting write-backs.
Throughout this work we focus on applying \gls*{vfi} in order to skip machine instructions.
In this context, we further define the \emph{Fault Target} as the machine instruction the adversary aims to skip, in order to cause a specific misbehavior.
Moreover, a Fault Target is assumed to be \emph{hit}, once it is successfully attacked by injecting a fault.

\vspace{-13pt}
\paragraph{Fault Injection Setups} can be divided into two main classes
of \emph{Cooperative} \gls*{fi} and \emph{Non-Cooperative} \gls*{fi}~\cite{cryptoeprint:2020:937}.
In the former, the attacker is able to reprogram the \gls*{dut}.
Here, the attacker commonly implements a protocol to communicate with the device.
By this it is possible to, e.g., call a subroutine on the device that shall be tested against \gls*{fi}, by issuing a corresponding command.
In addition, in cooperative \gls*{fi}, the \gls*{dut} is notifying the \gls*{fi} framework when entering the code region to be tested by asserting the synchronizing signal, referred to as \texttt{Trigger}.
Cooperative setups are commonly encountered in \glspl*{poc}.
In the latter the adversary is unable to reprogram the \gls*{dut}.
Non-Cooperative \gls*{fi} is commonly encountered in attack scenarios, which focus on attacking proprietary targets.

\section{Adversary Model}
\label{sec:advmodel}
The adversary model includes a physical access attacker. 
Further, the adversary is capable of performing slight modifications to the \gls*{dut} in order to make \gls*{vfi} possible, similar to related literature~\cite{o2016fault} (e.g., attaching copper wires, detaching bypass-capacitors).
In order to define Fault Targets, the adversary has some knowledge about the targets firmware, which may be, e.g., through the use of a public library~\cite{238594} or a previous binary firmware disclosure~\cite{10.1007/978-3-030-15462-2_12}. 

\section{\coolname Design}\glsresetall
\label{sec:attack_flow_mfi}

\begin{figure}
    \centering
    \includegraphics[width=0.35\textwidth, trim={12cm 4.7cm 13.2cm 7cm},clip]{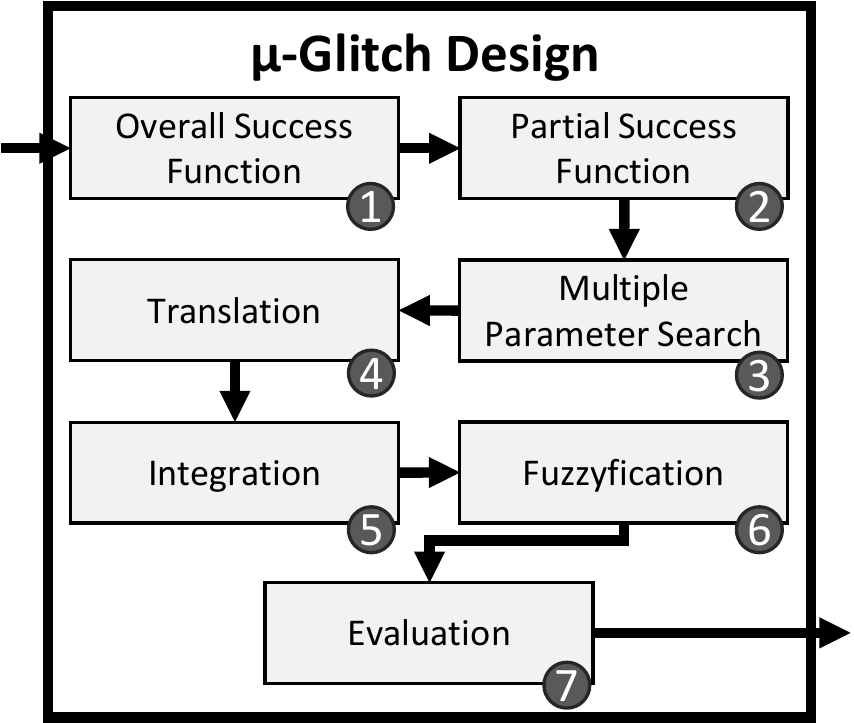}
    \caption{High level overview of our \coolname design}
    \label{fig:DESN}
\end{figure}

In this section, we present our novel \gls*{mfi} design, named \coolname.

Similarly to \gls*{sfi} attacks, we adopt the high-level flow, which consists of defining the experiment success function, performing fault injection by exhaustively searching the fault's parameters and analyzing and comparing success rates\footnote{\url{https://github.com/newaetech/chipwhisperer-jupyter/tree/92307484b155394a01c7021f1d21123efccee4aa/courses/fault101}}.

The complexity of searching multiple fault parameters at once increases exponentially by every additional fault, using conventional parameter search algorithms. Therefore, we introduce our novel, efficient \textit{sweeping} approach to be used in \gls*{mfi} setups.

The overall attack flow is depicted in \autoref{fig:DESN}:
\begin{enumerate}

    \item \emph{Define Overall Success Function}:\\
    In order to decide about the outcome (i.e. overall success or failure) of a  \gls*{mfi} attempt, a \gls*{sf} has to be defined.
    This is a binary function which is evaluated in later steps in order to indicate either a \texttt{success}, if and only if all the Fault Targets are hit at once 
    or a \texttt{failure} in all other cases.
    
    \item \emph{Define Partial Success Functions}:\\
    The parameter search needs to distinguish between hit Fault Targets. This is achieved by defining \glspl*{psf},
    which are needed to recognize, if some, but not all, Fault Targets have been hit, whereas the \gls*{sf} only allows to determine, if \emph{all} the Fault Targets have been faulted during a single, consecutive execution.
    
    \item \emph{Perform Multiple Parameter Search}:\\
    The goal of performing the Multiple Parameter Search is equal to this of the Parameter Search in \gls*{sfi}, i.e. valid fault parameters have to be discovered, which lead to an experiment \texttt{success}.
    Even though there are multiple such parameter pairs to be discovered throughout this step, the process of finding the right parameters is similar to the one in \gls*{sfi}:
    The adversary uses a \emph{single fault} in order to search in an increased space, spanned by all Fault Targets parameters.
    We refer to this process as \emph{sweeping}.
    As \glspl*{psf} have previously been defined, when the injected fault hits one of the Fault Targets, it will be detected by evaluating the corresponding \gls*{psf}.

    As there is only a single fault per execution injected, the \gls*{sf} is not evaluated during this step.
    The result is a set of sets of absolute parameters, i.e., one set per Fault Target, absolute to a common synchronization \texttt{Trigger}.
    
    \item \emph{Translating Absolute Parameters}:\\
    The previously discovered \emph{absolute} fault parameters need to be translated into relative parameters (i.e., relative to the preceding fault), by using the inductive definitions in \autoref{eq:first_local} and \autoref{eq:further_local}.
    \begin{equation}
        \label{eq:first_local}
        R_0 = A_0
    \end{equation}
    \begin{equation}
        \label{eq:further_local}
        R_n = A_n - \left(A_{n-1} + W_{n-1}\right)
    \end{equation}
    That is, the first faults relative \texttt{Offset} $R_0$ is always equal to the first absolute \texttt{Offset} $A_0$ found by our sweeping approach.
    Every additional fault's relative \texttt{Offset} $R_n$ is defined recursively in terms of its absolute \texttt{Offset} $A_n$, its previous global \texttt{Offset} $A_{n-1}$ and the previous fault's width $W_{n-1}$.
    All the fault's \texttt{Widths} may be directly adopted.
    
    \item \emph{Fuzzyfy Parameters}:\\
    Due to the non-deterministic behavior of the \gls*{dut} in the presence of voltage faults, every preceding fault may affect its succeeding ones parameters in unpredictable ways.
    In order to address this uncertainty, slight modifications have to be applied to the relative \texttt{Offsets}. 
    We refer to this process as \emph{fuzzyfication}.
    Here, every fault's \texttt{Offset} is not considered a single value, but rather a very small interval bound by $\pm\Psi: \Psi \in \field{N}$.
    The hereby generated intervals serve as input for the following integration step.

    \item \emph{Integrate Fuzzyfied Parameters}:\\
    As the uncertainty and hence the provided sets are considered to be very small, it is viable to perform an exhaustive search on multiple fault's parameters.
    As all the required voltage faults are hereby injected at once, this represents the first step in which the actual attack is performed.
    Therefore, each \gls*{mfi} attempt is evaluated based on the overall \gls*{sf}, instead of the \glspl*{psf}.
    
    \item \emph{Evaluate and Analyze Repeatability}:\\
    Finally, analyzing combinations of multiple fault's parameters, which in combination led to an overall \texttt{success} is needed.
    Further, if there are multiple such combinations available, the different combinations have to be qualified and compared w.r.t. their \texttt{success rates}.
    
\end{enumerate}

\subsection{Transforming Non-Cooperative Setups}\label{sec:transfer}
\coolname is also able to cope with non-cooperative setups, in which there are no \glspl*{psf} definable, by transforming non-cooperative setups to cooperative ones.
In general, an adversary can perform parameter search on a physical identical, but cooperative setup, before transferring the identified parameters to the non-cooperative setup, as in this case defining \glspl*{psf} is easy ~\cite{cryptoeprint:2021:1217}. 
This approach is feasible whenever the firmware can (partly) be reproduced and the offset in between Fault Targets remains constant.

As every \gls*{ic} manufacturer (e.g., STM, Atmel and NXP) provide their own \gls*{sdk} in order to generate the machine instructions for the target device, and it is considered to be best practice and highly encouraged by the manufacturers to use this example code to set up \gls*{tzm}, every non-cooperative setup using these public \glspl*{sdk} can be transformed to a cooperative one in the sense of the parameter search.

Therefore, in such a scenario it is possible to transfer the parameters of several Fault Targets to a non-cooperative setup, thus reducing complexity drastically and enable  attacking proprietary black-box software on these devices.

\section{\coolname Attack on TrustZone-M}\glsresetall
In this section we describe a real-world attack on \gls*{tzm} using \coolname, as compromising these highly secured environments usually leads to the disclosure of sensitive information.
The goal of our attack is to leak secrets stored in secure memory of the \gls*{tzm} from within non-secure firmware. 
Even though \coolname is able to circumvent all duplication based instruction-level countermeasures, as stated in \autoref{sec:related}, we chose to attack the duplication based protection of NXP.
NXP is an \gls*{ic} manufacturer, which recently adapted \gls*{fi} countermeasures similar to protections proposed by academia in their real-world threat modeling processes for their ARMv8-M series \glspl*{mcu}. The chips are protected by implementing a modified version of the duplication based approach of Barenghi et al.~\cite{10.1145/1873548.1873555}, which is referred to as \emph{Duplicate Registers}~\cite{lpc55s69, rt6xx}. 
We will elaborate this in more detail in \autoref{sec:dup_reg}. We also evaluate this attack on other ARMv8-M \glspl*{mcu} (namely, \texttt{STM32L5} and Atmel \texttt{SAML11}) which can be compromised with a subset of Fault Targets of this attack. First, we provide preliminary background on \gls*{tzm}.

\subsection{TrustZone-M Background}
\begin{figure}
    \centering
    \includegraphics[width=0.45\textwidth, trim={12cm 4.4cm 11.8cm 7cm},clip]{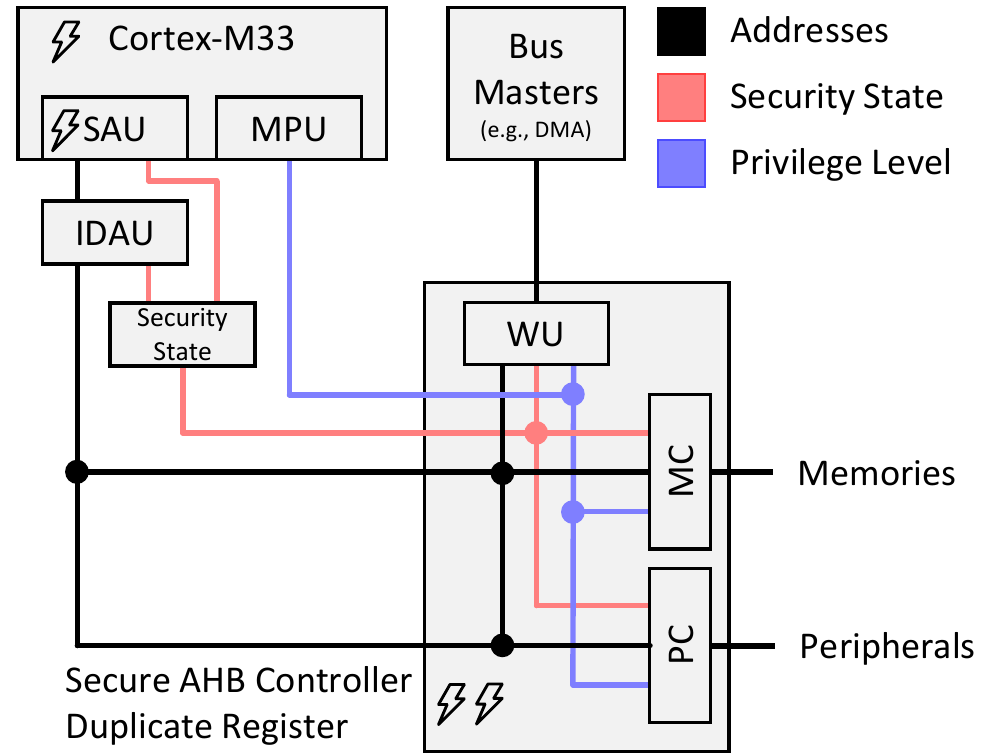}
    \caption{Example implementation of the Backchecking Mechanism on the \gls*{ahb}, a flash denotes parts to be faulted}
    \label{fig:SEC_AHB_CONTROLLER}
\end{figure}

\gls*{tzm} is a \gls*{tee} for embedded processors, i.e., it introduces system wide hardware-enforced computation and memory isolation mechanisms, which are built directly into the processor.
The \gls*{tzm} platform is configured by different memory attribution units, which are elaborated in the following and depicted in \autoref{fig:SEC_AHB_CONTROLLER}.
\vspace{-13pt}
\paragraph{The \gls*{sau}} is specified and designed by ARM.
Here, it is possible to define up to eight different memory regions which can be either Secure (\texttt{S}), Non-Secure (\texttt{NS}) or Non-Secure Callable (\texttt{NSC}).
The final security state of a memory region, is defined in conjunction with the \gls*{idau}.

\vspace{-13pt}
\glsreset{idau} 
\paragraph{The \gls*{idau}} is specified by the corresponding \gls*{ic} manufacturer.
It defines memory regions to be either \texttt{S} or \texttt{NS}. 
Defining \texttt{NSC} is a privilege granted exclusively to the \gls*{sau}.
On most of the commercially available ARMv8-M \gls*{tzm} processors the \gls*{idau} is implemented to perform a bit check on the $28_{th}$ bit of a requested address.
For an arbitrary address, the \gls*{idau} returns \texttt{S} if the $28_{th}$ bit is set, otherwise \texttt{NS}.

\vspace{-13pt}
\paragraph{The final Security State} is determined by the strongest output security state from \gls*{sau} and \gls*{idau} for a requested address, where the partial order of $S > NSC > NS$ holds.

\vspace{-13pt}
\paragraph{The Transition between \texttt{S} and \texttt{NS}}
\label{sec:bxns}
takes place based on novel ARMv8-M instructions.
The non-secure code is linked against a set of function headers, called the \emph{veneer table}, which is exported during compilation of the \texttt{S} firmware.
To switch from \texttt{NS} to \texttt{S}, it is required to take a detour through calling a veneer function in a \texttt{NSC} region, which consists of a Secure Gateway (\texttt{SG}) instruction\footnote{\url{https://developer.arm.com/documentation/100690/0201/Switching-between-Secure-and-Non-secure-states}}.

To switch from \texttt{S} to \texttt{NS}, it is necessary to use either 
the Branch with eXchange to Non-Secure state (\texttt{BXNS})
or the Branch with Link and eXchange to Non-Secure state (\texttt{BLXNS}) instruction.
Here it is important to note, that the security level transition will only happen, if the \gls*{lsb} of the \texttt{NS} address is unset\footnote{\url{https://developer.arm.com/documentation/100235/0004/the-cortex-m33-instruction-set/branch-and-control-instructions/bxns-and-blxns}}.

\vspace{-13pt}
\paragraph{The Backchecking Mechanism}
\label{sec:intro_ahb}
represents a \gls*{tee} protection available on \glspl*{mcu} featuring \gls*{tzm}, which ensures confidentiality and integrity on a system level.
In this concept, additional integrity checks are realized by introducing new hardware units on the internal system bus, referred to as \gls*{ahb}.

The three novel hardware units residing on the \gls*{ahb} matrix are:
\begin{description}
    \item[\glspl*{wu}] 
    are used to wrap \gls*{tzm} unaware bus masters to signal side-band information on the \gls*{ahb} matrix. This additional information determines the Security State and the Privilege Level for a requested address. Based on this information, the Checker Units perform additional checks upon every bus access.
    
    \item[\glspl*{mc}]
    are used to protect memory devices, such as Flashes, \glspl*{ram} and \glspl*{rom} from unintended access of an application.
    
    \item[\glspl*{pc}]
    are used to protect the peripherals which are directly connected to the \gls*{ahb} or via \gls*{apb} from unintended access of an application.
\end{description}
NXP refers to this concept as Secure \gls*{ahb} Controller~\cite{lpc55s69}, STMicroelectronics to \gls*{gtzc}~\cite{stm32l5} and Atmel to \gls*{pac}~\cite{saml11}.

\subsection{Attack Internals}
Throughout this section we describe our concrete attack against NXP based implementations of the TrustZone-M and their \gls*{fi} countermeasures.

\subsubsection{NXP's Duplicate Register}\label{sec:dup_reg}
NXP's \gls*{fi} countermeasure, referred to as the Duplicate Registers method, deploys for every security-critical register a second, equally structured register in memory space. 
If active, both of these registers are written sequentially in firmware.
Once \gls*{sfi} is applied in order to skip or modify an assignment to a secured register, it's duplicate register would afterwards still be written as intended.
Based on the introduced inconsistency between original register and it's duplicate register, any \gls*{sfi} attempt is detected in hardware.
This advanced countermeasure is, e.g., encountered in NXP's \texttt{LPC55S6X}~\cite{lpc55s69} and \texttt{RT6XX}~\cite{rt6xx} series \glspl*{mcu}.
We have identified Duplicate Registers being used in Debugging Features, \gls*{puf} Index Configuration and the Activation of the Secure \gls*{ahb} Controller.

\subsubsection{Interdependency of Protections}
As the checks performed by \gls*{sau} and \gls*{idau} and the checks performed by NXP's Secure \gls*{ahb} Controller are performed sequentially, attacking only one of these checks would always be detected by the respective counterpart.
In addition, the activation of the Secure \gls*{ahb} Controller is further protected by Duplicate Registers.
Hence, in order to succeed, all of these have to be successfully attacked during one consecutive execution.
While conducting our experiments, NXP processors seem to lock themselves into erroneous states, when the TrustZone-M specific instructions to switch the security context are issued, whenever the \gls*{sau} is not active.
Moreover, the activation of the \gls*{idau} cannot be skipped, as it is active by default after \gls*{por}.

\subsubsection{Fault Targets}
\label{sec:abstract_fault_targets}
With the \gls*{fi} countermeasure and interdependencies of the \gls*{tzm} protections in mind, we define the following Fault Targets and their technical details, which are represented by a flash symbol in \autoref{fig:SEC_AHB_CONTROLLER}.

\glsreset{sau}
\vspace{-13pt}
\paragraph{Activation of the \gls*{sau}}
\label{sec:technical_fault_targets}
The \gls*{sau} is activated in the \gls*{tzm} setup routine, a system routine that is executed before the trusted and secure user code is executed.
The relevant Fault Target is depicted in line 6 of \autoref{lst:init-trustzone}, and is commented with \texttt{SAU\_CTRL}.

\begin{figure}
    \begin{lstlisting}[
        style=sourcecode, 
        language=C, 
        caption=Simplified version of NXP's TrustZone-M Setup Routine, 
        label={lst:init-trustzone}]
BOARD_InitTrustZone:
    ; configure secure regions
    ...
    ; activate SAU
    MOVS R2, #1
    STR  R2, [0xE000EDD0]; SAU_CTRL
    ; configure secure AHB controller
    ...
    ; activate secure AHB controller 
    MOVW R2, #0xAAA5
    STR  R2, [0x500ACFF8]; Duplicate
    ...
    MOVW R2, #0xAAA5
    STR  R2, [0x500ACFFC]; Original
\end{lstlisting}
\end{figure}

Once the Fault Target is hit, i.e. the \texttt{STR} instruction is skipped, the \gls*{sau} is disabled.
As later on the Secure \gls*{ahb} Controller is setup as intended, it would detect any invalid bus accesses due to a mismatch of the configuration of both hardware units.
This inconsistency between the Secure \gls*{ahb} Controllers configuration and \gls*{tzm} configuration is what prevents a successful attack at this point.
\vspace{-13pt}
\paragraph{Activation of Secure \gls*{ahb} Controller}
To resolve this inconsistency, the adversary must prevent the Secure \gls*{ahb} Controller from being activated.
The Fault Target that needs to be hit in order to prevent activation is shown in line 15 of \autoref{lst:init-trustzone} and is commented with \emph{Original}.
Once this store instruction is skipped, the Secure \gls*{ahb} Controller is being kept deactivated.
Due to the use of \emph{Duplicate Registers} however, any successful \gls*{fi} attempt will still be detected.
\vspace{-13pt}
\paragraph{Duplicate Register for Secure \gls*{ahb} Controller}
At this point the processor would be able to detect an inconsistency between the deactivated Secure \gls*{ahb} Controller and its active Duplicate Register.
Hence, the assignment to the Duplicate Register forms another Fault Target, which is shown in line 13 of \autoref{lst:init-trustzone} and is commented with \emph{Duplicate}.
\vspace{-13pt}
\paragraph{Prevent Switching of Security Context}
\label{sec:privilege_escalation}
After the previous Fault Targets are all hit, the \gls*{sau} and all the \gls*{tzm} protections are fully disabled.

After the boot process, the bootloader passes execution to \texttt{S} firmware, in which the \gls*{tzm} is configured\footnote{\url{https://www.nxp.com/design/software/development-software/mcuxpresso-software-and-tools-/mcu-bootloader-for-nxp-microcontrollers:MCUBOOT?tab=Design\_Tools\_Tab}}.
Afterwards, the main function of the user defined code sets up the system and peripherals, before ultimately passing the execution to \texttt{NS} code, which at this point is invalid and therefore the \gls*{ic} would lock itself into an erroneous state. This transition is performed by the \gls*{tzm} specific \texttt{BXNS} instruction.
In \autoref{lst:security_state_transition}, we present the relevant disassembled parts of the binary, generated by \gls*{gcc} for transitioning from \texttt{S} to \texttt{NS}.

\begin{figure}
    \begin{lstlisting}[
    style=sourcecode,
        language=C, 
        caption={GCC generated Code to switch the Security Context},
        label={lst:security_state_transition}]
__gnu_cmse_nonsecure_call:
    ; clear all registers, prevent leaks
    ...
    ; transition to non secure world based
    ; on parameter R4
    BXNS    R4
; secure main function
main:
    ; Set up system and peripherals
    ; Execute user specific code
    ...
    ; Unset the LSB by shift-out-shift-in
    ; R4 is storing the NS destination address
    LSRS    R4, R4, #1
    LSLS    R4, R4, #1
    ; clear unbanked registers, prevent leaks
    ...
    ; transition to NS will be handled here
    BL      __gnu_cmse_nonsecure_call
    ; endless loop, should never be reached
    B .
\end{lstlisting}
\end{figure}

In order to prevent the internal security context switch from \texttt{S} to \texttt{NS}, it is sufficient to skip the shift-out shift-in operation, implemented by the \texttt{LSRS} and \texttt{LSLS} instructions defined in line 16 and 17.
As this code is executed in \texttt{S} state, the \gls*{lsb} is always set.

It is worth noting that even though we have been using NXP's toolchain to generate the firmware, the herein described privilege escalation can be performed on almost every commercial available \gls*{tzm} \gls*{mcu}, as most \gls*{ic} manufacturer release their toolchains based on \gls*{gcc}.
Moreover, in ARM based compilers clearing the \gls*{lsb} is performed by a single bit clear (\texttt{BIC}) instruction.
\subsection{\coolname Hardware Framework}
The Fault Targets in the \gls*{mfi} attack, need to be successfully hit \emph{all at once, in one continuous execution} in order to fully break the security granted by NXP's \gls*{tzm} implementation.
A lack of suitable commercial \gls*{mfi} tools led to the development of our \coolname \gls*{mfi} framework, which will be introduced throughout this section.

Our custom \gls*{mfi} framework consists of six components, namely the Clock Generation Unit,
the Host Communication Unit, the I/O Buffer Unit, internal Configuration Registers, the Multiple Voltage Fault Unit and the Serial Target I/O Unit.
In the following, we elaborate the Multiple Voltage Fault Unit in more detail, as it represents one of the main parts of our framework.
However, in order to introduce our \gls*{mfi} hardware, it is important to first understand the \gls*{sfu} design, which is commonly encountered in \gls*{sfi} setups\footnote{\url{https://github.com/chipfail/chipfail-glitcher/tree/master/chipfail-glitcher.srcs/sources\_1/new}}.

\begin{figure}
    \centering
    \includegraphics[width=0.45\textwidth, trim={12cm 7.5cm 11.8cm 7.5cm},clip]{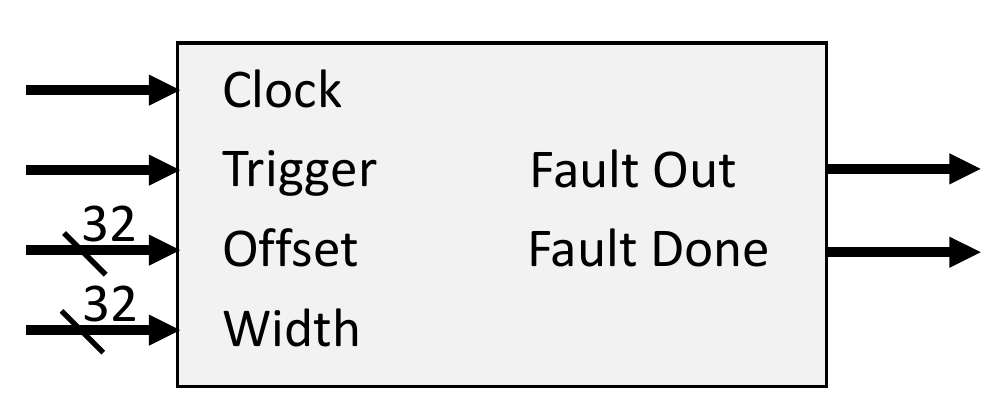}
    \caption{Block Diagram of a Single Fault Unit}
    \label{fig:single_glitch_unit}
\end{figure}

\autoref{fig:single_glitch_unit} shows a block diagram of a typical \gls*{sfu}.
Similar designs are commonly used to inject a single voltage fault into a \gls*{dut}.
A voltage fault is defined by its \texttt{Width} and its \texttt{Offset}, w.r.t. a synchronization point referred to as the \texttt{Trigger} (cf. \autoref{sec:voltage_fault}).
In our \gls*{sfu} design, the \texttt{Width} and \texttt{Offset} are both defined as $32-$Bit inputs.
A \gls*{sfi} is initiated, once the single bit \texttt{Trigger} input signal is asserted.
Starting with this external \texttt{Trigger} event, the hardware starts counting, using the reference \texttt{Clock} signal.
Once the defined \texttt{Offset} has been reached, the \texttt{Fault Out} signal is set high and remains high for exactly \texttt{Width} clock cycles.
The \texttt{Fault Out} signal is routed directly into the gate of a N-Channel \gls*{mosfet}, with its source connected to the fault voltage level and its drain connected to the power supply line of the \gls*{dut}.
This circuit is referred to as the \emph{Crowbar-Circuit}~\cite{o2016fault} and represents the state-of-the-art method used to inject a voltage fault.
To indicate that the fault attempt has been processed, an additional output signal named \texttt{Fault Done} is asserted for a single clock cycle.
\subsubsection{Multiple Fault Injection Hardware}
The \gls*{sfu} discussed previously builds the base for the design of our \gls*{mfi} unit.
By chaining multiple \glspl*{sfu} together, we are able to inject multiple, coordinated faults using a single trigger.
For this purpose, the \texttt{Trigger} input of a unit is connected to the \texttt{Fault Done} signal of its predecessor, whereas the first \gls*{sfu}'s \texttt{Trigger} is directly connected to the external \texttt{Trigger} signal, forming a chain of units.
The \texttt{Fault Out} lines are combined by using the logical \texttt{Or} operator, whereas the signal indicating the termination of each \gls*{mfi} attempt is solely defined by the last \gls*{sfu}'s \texttt{Fault Done} signal.
By chaining \glspl*{sfu} a single trigger signal can be used to perform \gls*{mfi}.

\begin{figure}
    \centering
    \includegraphics[width=0.45\textwidth, trim={13.5cm 4.4cm 13.5cm 6.5cm},clip]{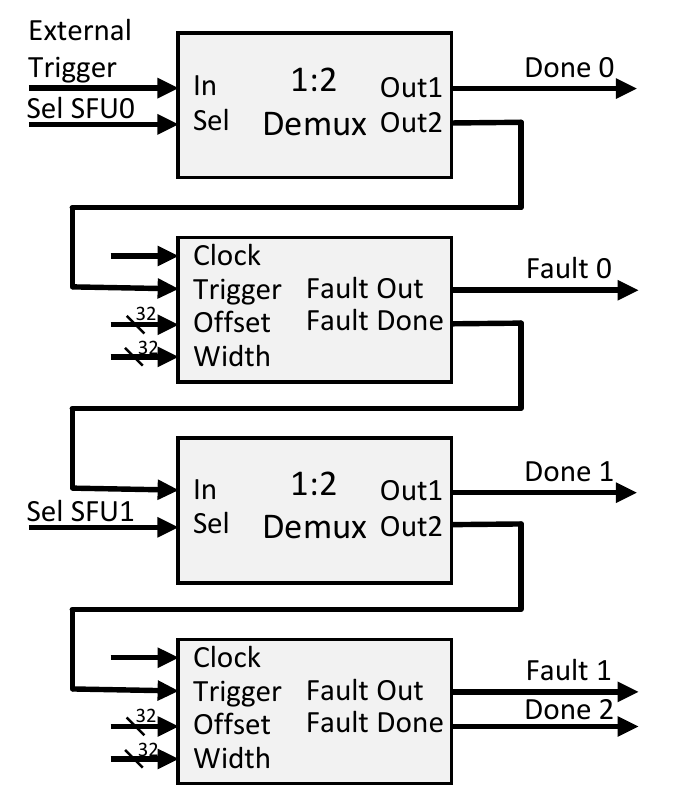}
    \caption{Block Diagram of the Multiple Voltage Fault Unit}
    \label{fig:MFI_unit}
\end{figure}

While this approach is capable of injecting multiple, coordinated faults into a \gls*{dut}, the number of injected faults is fixed by design.
Once there are $n \in \field{N}$ \glspl*{sfu} downloaded to the \gls*{fpga}, the framework is determined to generate exactly $n$ voltage faults, each time it gets triggered.
With respect to our attack flow described in \autoref{sec:attack_flow_mfi}, the dynamical configuration of the number of injected faults is desirable.
Hence, we propose a slightly more complex design, which enables the dynamic activation and deactivation of \glspl*{sfu}, even after the hardware has been downloaded to the \gls*{fpga}. By introducing multiplexers in between each of the \glspl*{sfu}, as shown in \autoref{fig:MFI_unit}, the design is dynamically configurable.

In this depiction, the routing of the \texttt{Clock}, \texttt{Offset} and \texttt{Width} signals of each \gls*{sfu} are intentionally omitted, in order to focus on the interconnection of multiple \glspl*{sfu}.
In addition, all the \texttt{Fault Out} signals (\texttt{Fault 0}, \texttt{Fault 1}, \ldots) are assumed to be combined using the logical \texttt{Or} operator.
The hereby generated signal controls the gate of the \gls*{mosfet} of our Crowbar circuit implementation.
The blocks labeled \texttt{1:2 Demux} are demultiplexers, i.e., hardware units which forward an input signal (\texttt{In}) to one of multiple output signals (\texttt{Out1} or \texttt{Out2}), depending on the state of another input signal (\texttt{Sel}).
The selection lines (\texttt{Sel SFU0}, \texttt{Sel SFU1}, \ldots) may be modified by the controlling host system, by updating the \gls*{mfi} framework's internal registers.
The first demultiplexer, which routes its input to \texttt{Out1} asserts one of the \emph{Done} signals, which indicates the end of this \gls*{mfi} attempt and interrupts the forwarding of \texttt{Done}-Signals to succeeding \glspl*{sfu}.
Each of the \texttt{Fault Done} signals (i.e. \texttt{Done 0}, \texttt{Done 1}, \texttt{Done 2}, \ldots) are also combined using the logical \texttt{Or} operator, to form a single output signal.

\section{Evaluation}\glsresetall
\label{sec:evaluation}
To thoroughly test our \coolname \gls*{mfi} hardware design and novel parameter search algorithm, we conduct multiple evaluations throughout this section.
First, we show the feasibility of attacking the Duplicate Registers countermeasure, this method is mimicked and attacked based on a cooperative as well as a non-cooperative firmware simulation.
By this our novel approach of searching for parameters can be compared to performing an exhaustive search, with optimized conventional parameter search algorithms.
Moreover, in order to prevent the security context switch, we perform a privilege escalation on code generated by \gls*{gcc}.
The Fault Target are two immediately successive instructions, which have both to be attacked at once.
For this purpose we evaluate, whether it is more promising to inject two, narrow faults or a single, wider fault when aiming at attacking such successive instructions.
We conclude our evaluation by performing our real-world attack, in which we attack NXP's \gls*{tzm} implementation with all the new protections activated, by hitting all Fault Targets introduced in \autoref{sec:abstract_fault_targets} in one execution.

\subsection{Attacking Duplicate Registers}
\label{sec:simulation_duplicate_registers}
To show the feasibility of circumventing the Duplicate Register \gls*{fi} countermeasure and evaluate our proposed \gls*{mfi} parameter search algorithm 
we implement an isolated simulation of this countermeasure.
It is possible without loss of generality, since the Duplicate Register method and our simulation both translate to equal Fault Targets.
In the following we describe our simulation approach for cooperative as well as non-cooperative parameter searches.
Afterwards, multiple parameter searches for both the different setups are performed.
We conclude this simulation by analyzing the success rates for attacking our simulation of the Duplicate Registers \gls*{fi} countermeasure.

\subsubsection{Fault Targets}
In order to simulate the Duplicate Registers Fault Injection countermeasure, we first define two functions, which are used for the different \gls*{mfi} setups (i.e., cooperative and non-cooperative), and whose source code is provided as a reference in the Appendix in \autoref{lst:parameter_non_cooperative} and \autoref{lst:cooperative_fault_target}.
In both functions two, zero initialized variables are defined.
These represent our Fault Targets, i.e., a register to be protected and its duplicate register.
After their definitions, these are written sequentially with the same, non-zero value.
The goal of the adversary is to skip both the assignments in one execution, hence compromising the protection introduced by the Duplicate Registers method.
In our cooperative simulation \glspl*{psf} are defined, which allows the use of our efficient approach, as described in \autoref{sec:attack_flow_mfi}.
For comparison, in the non-cooperative simulation, \glspl*{psf} are assumed to be impossible to define, and an optimized exhaustive search is performed. We chose to use the exhaustive search approach as proposed by NewAE, as it represents the state-of-the-art in \gls*{sfi} attacks\footnote{\url{https://github.com/newaetech/chipwhisperer-jupyter/tree/92307484b155394a01c7021f1d21123efccee4aa/courses/fault101}, which tests all possible combinations of \gls*{fi} parameters. This search requires no additional knowledge in contrast to other approaches which require time consuming model training~\cite{wu2020fast} for every glitch or the definition of a fitness function~\cite{10.1007/978-3-319-08302-5_16}, which may not always be possible.}
Each assignment to a Fault Target is preceded with a random delay, which is determined during compile-time.
Through this, the binary firmware exhibits exactly the same behavior during each execution.
Once the firmware is re-compiled, another binary is generated which exhibits a different timely behavior.
This way we ensure that parameters discovered throughout one experiment are very unlikely to fit another experiment.
In order to compare the different setups, the same delays for both setups have been used.

\subsubsection{Parameter Search}
The different search algorithms, used to discover \gls*{mfi} parameters, are performed several times based on the introduced random delays, before comparing the results of the different approaches to each other.
\autoref{table:comparison_glitch_search} compares an exhaustive search (left) to our efficient sweeping approach (right).
Each row represents a single comparison of an exhaustive search to our cooperative approach, based on the depicted compile-time delays.
The results for applying our sweeping approach are divided into the time to search the parameters based on \glspl*{psf} (left) and their integration (right).
It shows that our novel approach in a \gls*{mfi} context is much more efficient than the traditional exhaustive search.
Here, for every comparison our novel approach required approximately two percent of the time required by the exhaustive search.

\begin{table}
    \centering
    \begin{tabular}{|l|l|l|l|l|}
    \hline
    Delay 1 & Delay 2 & Exhaustive  & Sweeping              \\ \hline
    7       & 43      & 12:31h      & 00:06h + 00:09h       \\ \hline
    33      & 19      & 21:01h      & 00:18h + 00:10h       \\ \hline
    4       & 50      & 07:46h      & 00:02h + 00:09h       \\ \hline
    22      & 1       & 16:04h      & 00:11h + 00:10h       \\ \hline
    \end{tabular}
    \caption{Comparing optimized, exhaustive search (Exhaustive) to our efficient approach (Sweeping)}
    \label{table:comparison_glitch_search}
\end{table}

\subsubsection{Evaluation And Repeatability}
For each of the previous \gls*{mfi} parameter searches, multiple combinations of \gls*{mfi} parameters have been returned, which have shown to evaluate the \gls*{sf} to \texttt{success}.
Moreover, both parameter search algorithms yielded similar fault parameters, which exhibit only negligible differences.
When performing an actual attack, an adversary would always choose the combination of parameters, which has the highest probability for an attack to succeed.
Therefore, each previously determined successful combination has to be qualified by means of their reliability and repeatability.

For this we first define the \emph{most promising attack parameters} as the parameters, which show the highest success rate of our attack.
To estimate the most promising parameters, we perform \gls*{mfi} $1,000$ times based on different successful parameter configurations.
Once the most promising attack parameters have been determined, $100,000$ \gls*{mfi} attempts are performed.
The experiment returned a success rate of $0.212$ and a respective failure rate of $0.788$.
This means, that by using our highly precise \gls*{mfi} design about every fifth attempt of injecting two consecutive faults into our \gls*{dut} in order to overcome the Duplicate Registers simulation succeeded.

\subsection{Attacking Successive Instructions}
\label{sec:successive_instructions}
As mentioned in \autoref{sec:abstract_fault_targets}, besides skipping the activation of the \gls*{sau} and the activation of the Secure \gls*{ahb} Controller, an additional privilege escalation is required.
The Fault Targets are two, directly succeeding shift instructions, which, under normal circumstances, unset the \gls*{lsb} of the \texttt{NS} target address.

With the following simulation, we aim to evaluate if immediately successive instructions, as they are encountered in this scenario, are best to be attacked by a single, wider fault or by multiple, narrow ones.

\subsubsection{Fault Targets}
Our Fault Targets are depicted in line 10 and 11 of our cooperative firmware example in \autoref{lst:succeeding_fault_targets}.
This simulation defines a variable labeled $a$ and assigns it an odd value of $0x13$, i.e., the \gls*{lsb} is set.
Based on the returned value, an adversary is able to distinguish which Fault Targets have been hit.
If, e.g., \emph{only the left shift} instruction has been skipped, then the returned value must be equal to $0x9$.
Under normal circumstances both shift operations are performed, clearing the \gls*{lsb} and resulting in a returned value of $0x12$.

\begin{figure}
    \begin{lstlisting}[
        language=C,
        style=sourcecode,
        caption={Simulation of our privilege escalation. The inline assembler performs the shift-out-shift-in operation used by \gls*{gcc} to clear the \gls*{lsb} of the NS target address},
        label={lst:succeeding_fault_targets}]
#include <stdint.h>
uint32_t succeeding_fault_targets(void) {
    uint32_t a = 0x13;
    set_trigger()  ; // synchronization
    reset_trigger(); // purposes
    asm volatile (
     "lsrs%[address],%[address],#1" "\n\t"
     "lsls%[address],%[address],#1" "\n\t"
     : [address]"=r" (a)
    );
    return a;
}
\end{lstlisting}
\end{figure}

\subsubsection{Parameter Search}
Throughout this simulation, we assume that the most promising attack parameters are already known, computed by using our efficient search approach.

\subsubsection{Evaluation And Repeatability}
We evaluate performing both, a broad \gls*{sfi} as well as two narrow \gls*{mfi} on the previously introduced Fault Targets to demonstrate how \gls*{mfi} compares to \gls*{sfi} when attempting to attack instructions that immediately follow another.
For this evaluation, $100,000$ \gls*{fi} attempts have been performed. 
Throughout this experiment we define the group of \texttt{invalid} results as these results, where the \gls*{dut} is either not responding after performing \gls*{mfi} or which cannot be explained by either skipping of the shift instructions.

Using a single, wider fault in order to fault two successive shift instruction resulted in a success rate of $24\%$, whereas attacking the same instructions using two, narrow faults resulted in a success rate of $15\%$. Moreover, the group of invalid results increased by $9\%$.
Hence, with respect to the success rate it is more reasonable for an adversary to inject a single fault, utilizing an increased \texttt{Width}, in order to attack the two successive shift instructions. The complete results are shown in the Appendix in \autoref{table:results_succeeding_fault_targets}.

\subsection{Attacking The TrustZone-M}
In this part, we evaluate our novel approach by attacking NXP's implementation of the \gls*{tzm} and all of the additional countermeasures in a real-world scenario.
Throughout this experiment, the \gls*{mfi} framework's frequency is specified as being $20$ times higher, than the frequency of the \gls*{dut}.
By oversampling of our \gls*{mfi} framework w.r.t. the \gls*{dut}, we gain additional precision for each of our \gls*{fi} attempts. Note, that it is possible to increase the success rate by increasing the oversampling rate or synchronizing the glitch to the \gls*{dut}'s clock signal\cite{o2014chipwhisperer}. This, unfortunately, would also mean to introduce more assumptions.

The firmware has been built based on the unmodified NXP's \gls*{sdk} \gls*{tzm} examples.
The Fault Targets have been described in detail throughout \autoref{sec:abstract_fault_targets}, as the activation of the \gls*{sau}, the activation of the Secure \gls*{ahb} Controller, the duplicate register for the activation of Secure \gls*{ahb} Controller and performing privilege escalation.

After all these Fault Targets have been hit during a single, consecutive execution of the targets firmware, the \texttt{NS} code is able to arbitrarily access any \texttt{S} and \texttt{NS} resources.
In the context of \glspl*{tee}, this represents a full compromise, as e.g., any secrets stored within secure memory may be disclosed and any secure defined peripheral may be arbitrary accessed and controlled from within non-secure code.

Moreover, up to this point all parts of our attack have already been proven feasible and practical, by attacking isolated simulations.
We are presenting our results, in terms of repeatability and reliability of our attack, in comparison with the conventional exhaustive parameter search.

\subsubsection{Search For Parameters}
The search for parameters is evaluated for a cooperative setup as well as a non-cooperative setup, in which no \glspl*{psf} can be defined.
The results of comparing both parameter searches are depicted in \autoref{tab:comparison_glitch_search_real}.

\begin{table}[]
    \centering
    \begin{tabular}{@{}c|c@{}}
        \hline
        Brute Force Search Time & 
        \begin{tabular}[c]{@{}l@{}} 
            Sweeping Search Time\\ (Search + Integration)
        \end{tabular} \\ \hline
        \textgreater 24h & 06:17h + 00:28h\\
        \textgreater 24h & 08:07h + 00:28h\\
        \textgreater 24h & 13:22h + 00:30h\\
        23:40h & 02:58h + 00:28h \\
    \end{tabular}
    \caption{Comparison of exhaustive search (left) to our novel sweeping approach (right) when searching the parameters of four voltage faults used to attack the TrustZone-M.}
    \label{tab:comparison_glitch_search_real}
\end{table}

Four different search passes for the same four Fault Targets have been performed with a limit of $24$ hours.
While the exhaustive approach led in only one of four parameter searches to a result within the given time limit, our sweeping approach has returned correct parameters for every single attempt.
Moreover, the fastest exhaustive search has shown to be eight times slower, than the corresponding sweeping approach.
The results of the sweeping approach are again split into the time it took to search the single fault parameters (left) and the time it took to integrate the respective parameters (right).
It is worth noting, that there is quite some variance contained in the depicted results, which can be explained by the non-deterministic behavior of \gls*{fi}, i.e., even if the parameters are perfectly set, the injected fault can never guarantee a success. 
Due to the non-deterministic discovery of fault parameters, we implemented our parameter searches to restart themselves, if either no total success has been observed for a non-cooperative setup, or not at least one partial success has been observed for every Fault Target, in a cooperative setup.

Note, that in this attack scenario it is possible to escalate a non-cooperative setup to a cooperative by using the parameter transfer described in \autoref{sec:transfer}.

\subsubsection{Evaluation and Repeatability}
After conducting the \gls*{mfi} parameter searches, for each setup a set of sets of fault parameter configurations has been returned, based on which the \gls*{sf} evaluated to \texttt{success}.
As the parameter found by both search algorithms differ only negligibly we used the parameters returned by our efficient, \gls*{mfi} parameter search, in order to determine the most promising attack parameters.
After these parameters are estimated, these are used to perform $1,000,000$ \gls*{mfi} attempts, in order to compromise the \gls*{tzm}.
Moreover, this evaluation has been performed two times.
The averaged results of both experiments are depicted in \autoref{table:tzm_res}.

\begin{table}[h!]
    \centering
    \begin{tabular}{lc}
        \hline
        Fault Targets     & Success Rate \\
        \hline
        SAU & $0.451$ \\
        SAU \& AHB CTRL & $0.0251$ \\
        SAU \& AHB CTRL \& DUPL & $0.0023$ \\
        SAU \& AHB CTRL \& DUPL \& PE & $0.0000003$ \\
        \hline
    \end{tabular}
    \caption{Results of performing our MFI attack against NXP's TrustZone-M.
    Results are given for an increasing amount of consecutive hit Fault Targets in one single execution}.
    \label{table:tzm_res}
\end{table}

The success rates when injecting exactly one (only disabling the \gls*{sau}) up to four (completely disabling \gls*{tzm}) faults based on the most promising attack parameters are depicted. 
\texttt{SAU} is referring to the first Fault Target, i.e., the activation of the \gls*{sau}, \texttt{AHB CTRL} is referring to the activation of the Secure \gls*{ahb} Controller, \texttt{DUPL} is referring to the Duplicate Register of the Secure \gls*{ahb} Controller and \texttt{PE} is referring to the privilege escalation. 
These result indicate that almost every second attempt ($45.1\%$) to inject four faults into the \gls*{dut} is deactivating the \gls*{sau}.
Moreover, $2.52\%$ of the performed \gls*{fi} attempts successfully disabled the Secure \gls*{ahb} Controller in addition.
An average of $0.23\%$ of \gls*{fi} attempts succeeded in attacking the activation of the \gls*{sau}, the activation of the Secure \gls*{ahb} Controller as well as the Duplicate Register of the Secure \gls*{ahb} Controller all at once.
And finally, $0.0003\%$ \gls*{mfi} attempts resulted in a total success.
Conducting such high amounts of \gls*{fi} attempts may seem excessive at first, however, performing one million \gls*{fi} attempts took only one and a half days, translating to one successful attempt in half a day.
With respect to the potential damage this practical attack may cause, once successful, we consider this attack critical.

\subsubsection{\coolname Transferability}

We evaluated the end-to-end attack for disabling \gls*{tzm}, both on cooperative as well as non-cooperative \gls*{mvfi} setups, on different target \glspl*{ic}, namely STMs \texttt{STM32L5}, Atmels \texttt{SAML11} and NXPs \texttt{LPC55S69} and \texttt{RT6600} \glspl*{mcu}.
In the Appendix in \autoref{tab:chips} we show the attacked chips and the necessary Fault Targets to achieve disabling \gls*{tzm} per chip. In addition, we show the time for baseline exhaustive search (capped at 48h) in contrast to our sweeping approach and the combined success rate of all glitches combined. \texttt{RT600} is conceptionally similar to the \texttt{LPC55S69} and can be attacked using the same Fault Targets. \texttt{SAML11} configuration is stored in Non-Volatile-Memory (NVM) space and either modifying read information or glitching the bootloader is necessary. In order to disable the \gls*{bod} of the \texttt{STM32L5}, which may interfere with \gls*{vfi}, glitching of the \gls*{pll} configuration is necessary, in order to run the chip with a clock frequency of $\le 32MHz$ in case the chip is configured to run faster. According to our evaluation \coolname can successfully attack conceptionally similar \glspl*{ic}.

\section{Potential Countermeasure}\glsresetall
Inspired by the insights of our evaluation we propose a potential countermeasure against \gls*{mfi} attacks.
We propose an \gls*{ilc}, as this kind of countermeasure may also be applied on already deployed hardware, i.e., no novel hardware revision is required in order to deploy physical sensors.

By attacking the Duplicate Registers method throughout this work, we have shown that single fault injection protections on the instruction level may be overcome by injecting additional faults.
This is possible due to the fact that the Fault Targets are located at always the same offsets.
Hence, an adversary is able to conduct a multiple parameter search and thus form a reliable attack.

Our concept of \emph{randomizing the Duplicate Registers} makes use of random delays, in order to strengthen the Duplicate Register method against \gls*{mfi} attacks, by removing the possibility to search for \gls*{mfi} parameters.
For this, we propose the use of a compiler pass, which introduces a compiler attribute that can be used to security critical assignments.
As the random delay is only introduced when the aforementioned attribute is encountered, the overhead on the computation in general is negligible.

The goal is to generate machine instructions, which force the processor to stall for a small, random period of time.
With this, a parameter search can not be successfully carried out as the \texttt{Offset} of the Fault Targets varies on a per-execution base.
Hereby however, a trade-off is introduced between increasing the parameter search space by introducing a large delay and the minimization of overhead in processor time by using a small delay.
When conducting our experiments, the \gls*{mfi} frameworks internal frequency has been chosen to be up to $20$ times higher, than that of the \gls*{dut}.
Here we have observed, that a timing difference of less than a single \gls*{dut}'s clock cycle leads to no successes.
Hence we assume that the stall time can be kept low without compromising the security.

As by this method there is no possibility to conduct a parameter search, the best possible attack is to inject, random parameterized faults into the \gls*{dut}, which leads to impracticability of a \gls*{mfi} attack, by decreasing its success rate tremendously.

By utilizing this approach to attack two Fault Targets, which have a random preceding delay of 0-9 cycles, the probability of injecting two successful glitches is 100 times lower than without this protection.

\section{Discussion}
We argue \coolname to be applicable to most \glspl*{mcu} with \gls*{tzm}, as the adversary only needs to have access to the power supply of the IC, no need to supply the DuT‘s clock signal by using oversampling and the \gls*{tzm} setup code of most \glspl*{mcu} is open source knowledge in form of \glspl*{sdk}.

Naturally, \coolname is not limited to conduct attacks against \gls*{tzm}. Van den Herrewegen et al.~\cite{van2021fill} used \gls*{sfi} to deactivate debugging protections to exfiltrate data. In recent \glspl*{mcu} by NXP, the debug interfaces are protected by duplication based approaches, therefore, \gls*{sfi} attacks cannot be used anymore. Hence, in order to overcome the debug features protection, \coolname \gls*{mvfi} approach becomes mandatory.

We also evaluated using \coolname to overcome the mitigating effects of \gls*{bod} against \gls*{vfi}. Even though it is indented to be a safety feature to power down an embedded device whenever the battery based voltage supply drops below a certain threshold, it has also been shown to detect the voltage drops caused by \gls*{vfi}. In sampling based \gls*{bod} approaches the supply voltage is measured periodically, which is commonly encoutered in embedded devices, as it has a relatively low power consumption. If the \gls*{bod} sampling frequency is high enough, it may happen that a voltage fault can be detected. We therefore propose to split a single, wider voltage fault into several, narrower ones exhibiting a similar effect on the target. Based on this, we were able to overcome a
sampling based \gls*{bod} by using two voltage faults, instead of a single fault 
which triggered the BOD. Unfortunately, the success rate of using two narrower glitches instead of one, is lower, as the chance to find a sweet-spot is decreasing which each additional glitch. Exemplary, the original and the first of the split glitches have the same offset. The width of the original glitch is $400 ns$. 
A double-fault inheriting a similar experiment outcome is represented by the two widths of $170 ns$ and $140 ns$ with an in-between offset of $100ns$. A graphic depiction can be seen in the Appendix in \autoref{fig:splitglitch}.

\section{Related Work}\glsresetall\label{sec:related}
In the following, we provide a summary of existing \gls*{fi} attacks,
attacks on \glspl*{tee} as well as \gls*{fi} countermeasures.
\vspace{-13pt}
\paragraph{Fault Injection Attacks:}
Several \gls*{fi} attacks have been proposed over the last years, which we utilize in this work. Our work builds on top of Roth~\cite{123698251376}, who has attacked several implementations of the TrustZone-M by injecting a single voltage fault.
The author aimed at hitting Fault Targets, which set the lower bounds of \texttt{NS} regions, therefore extending the regions to ultimately accessing sensitive, secure data. For this purpose firmware examples were attacked, which did not activate the vendor specific backchecking mechanisms of TrustZone-M implementations. In contrast to this, we focus on providing a reliable and repeatable procedure of conducting \gls*{mfi} attacks, which is able to circumvent all protections which are enabled by default in NXP's \gls*{sdk}. Trichina et al.~\cite{5577278} propose two-fault attacks on protected CRT-RSA implementations running on an advanced 32 bit ARM Cortex M3 core. The authors performed two-fault \gls*{lfi} on a protected cryptographic application, as \gls*{lfi} exhibits a high spatial resolution. Nashimoto et al.~\cite{Nashimoto2017} combined stack based \gls*{bo} attacks with two-fault \gls*{cfi}, in order to prevent the \gls*{bo} from being detected. Colin O'Flynn~\cite{o2016fault} showed that \gls*{vfi} may exhibit a high timely resolution based on his proposed \emph{Crowbar} circuit, which we utilize in this work. We have shown that by using this circuitry it is further possible to inject multiple voltage faults within a short period of time. Bozzato et al.~\cite{Bozzato_Focardi_Palmarini_2019} replaced the \emph{Crowbar} circuit by a \gls*{dac}.
This way improving \gls*{vfi} by increasing transferability to other hardware, being able to attack \gls*{bod} enabled \glspl*{ic} and injecting faults based on arbitrary waveforms.
While the authors mention to be able to inject multiple faults, they fell short of showing \gls*{mfi} attacks using their hardware and provided no evaluation, as they focus in their work on increasing the reliability and repeatability of \gls*{sfi}. Timmers et al.~\cite{7774479} performed \gls*{vfi} in order to corrupt the instruction decoding stages of the internal processor pipeline, with the goal of hijacking the control flow by setting the \gls*{pc} to predefined addresses stored in general purpose registers.
In a later work, Timmers et al.~\cite{8167704} showed that in an embedded Linux \gls*{os} the privileges can be escalated from user to system privileges by performing \gls*{vfi}.
By performing \gls*{vfi} against AMD's \gls*{sp}, Buhren et al.~\cite{10.1145/3460120.3484779} where able to control the key management and by this, compromise the security of AMD's \gls*{sev}. An overview of fault attacks on embedded devices is provided by Yuce et al.~\cite{yuce2018fault}. Werner et al.~\cite{9237302} generate fault models for \gls*{lfi} based on fault injection simulations.

The first \gls*{mcfi} has been demonstrated by Blömer et al.~\cite{6976638}, attacking two consecutive instructions during a single execution by directly modifying the clock signal. 
For this attack to succeed the external clock signal has to be fed directly into the processing part of the \gls*{dut}. However, as of today, most \glspl*{ic} use \glspl*{pll} which were shown to gracefully protect against clock glitching. 
Further, \gls*{mfi} has been performed by Colombier et al.~\cite{10.1007/978-3-030-97348-3_9} using a \gls*{lfi} setup. Due to its sophisticated spatial and timely resolution, optical fault injection forms a promising candidate for \gls*{mfi}. \gls*{lfi} setups, however are quite costly. A single \gls*{lfi} setup is commonly encountered in the magnitude of \$100.000, whereas the cost for a \emph{multiple} \gls*{lfi} setup is even higher. Moreover, in order to conduct \gls*{lfi} attacks, in general a much more invasive preprocessing of the \gls*{dut} is required, in comparison to \gls*{vfi}. Electromagnetic Fault Injection is commonly not considered to be used in MFI setups, as the internal capacitor banks take too much time in order to be recharged, thus rapid successfully injected consecutive faults cannot be guaranteed.   

Commercial equipment for \gls*{mfi}, e.g., Alphanov's double laser fault injection microscope (D-LMS) and Riscure's VC Glitcher are fairly expensive and only shown to conduct \gls*{mlfi}. Devices from NewAE~\cite{o2014chipwhisperer} cannot conduct \gls*{mvfi} but need a separate trigger for every glitch to be injected.

In contrast, \coolname, is capable of reliably performing \gls*{mvfi} based on low-cost hardware. We aim at attacking setups, in which commonly only a single, synchronizing \texttt{trigger} signal can be asserted. 
To the best of our knowledge, the possibility of conducting \gls*{mvfi} has not been studied before.

\vspace{-13pt}
\paragraph{Cryptographic Attacks:}
\gls*{dfa} has first been described by Biham and Shamir~\cite{10.1007/BFb0052259}. 
It poses a cryptanalytic attack which exploits computational errors in order to disclose cryptographic keys.
In recent years publications attacking today's
\gls*{aes}~\cite{10.1007/978-3-540-45203-4_23, 10.1007/978-3-642-21040-2_15, soleimany2021practical}, \gls*{rsa} \cite{aumuller2002fault}, \gls*{des} \cite{wu2020fast}, recent cryptographic Hash Functions~\cite{7774477, maldini2018genetic} and many more~\cite{LI20083727, 10.1007/11502760_24, 10.1007/978-3-540-77048-0_22, 10.1007/978-3-540-71039-4_10} emerged.
\vspace{-13pt}
\paragraph{Trusted Execution Environment Attacks:}
In recent years attacks utilizing different attack vectors against popular \glspl*{tee} like ARM's\gls*{tz} as well as Intel's \gls*{sgx} were published.

Tang et al.~\cite{203864} presented the CLKSCREW attack, which exploited an on-chip energy regulation mechanism in order to break the security promises by ARM's TrustZone.
Kenjar et al.~\cite{251590} described another software controlled, but hardware based fault injection approach, in which the authors were able to compromise any operating mode of Intel processors by modifying the frequency and voltage through privileged software interfaces.
The authors showed that software management interfaces can be exploited to undermine the system's security. Qui et al.~\cite{10.1145/3427384.3427394} performed a software based voltage fault injection, by abusing the \gls*{dvfs} techniques for energy efficiency, allowing them to attack a secure software implementation of \gls*{aes}. Ryan~\cite{10.1145/3319535.3354197} showed, that ARM's \gls*{tz} is susceptible to cache based attacks, which exhibit high temporal precision, high spatial precision and low noise. The author was able to fully recover a 256-bit private key from Qualcomm's version of the hardware-backed keystore. Ning et al.~\cite{ning2019understanding} exploited security vulnerabilities in ARMs software debugging features to extract sensitive information from \gls*{tz}. Jang et al.~\cite{10.1145/3152701.3152709} performed a denial-of-service against \gls*{sgx}, in which the CPU could be shutdown by performing a Rowhammer~\cite{8708249} attack. Lee et al.~\cite{203696} found that it is indeed possible to circumvent the hardware protections provided by the \gls*{sgx} design by performing \gls*{rop} attacks.

Different attack vectors were used in the past to attack TEEs, including \gls*{tz}, however these attack vectors are not suitable for \gls*{mvfi} and therefore not able to overcome \gls*{sfi} based protection mechanisms such as utilized by \texttt{LPC55SXX} and \texttt{RT6XX} \glspl*{mcu}.
\vspace{-13pt}
\paragraph{Fault Injection Countermeasures:}
Due to the vast amount of proposed work in this field, we present an overview in the Appendix in \autoref{tab:countermeasures}, where we list several proposed \gls*{fi} countermeasures, which are further classified into  \glspl*{ilc}~\cite{lpc55s69, rt6xx, 10.1145/1873548.1873555, Sakamoto2021, 2014, 10.1145/2858930.2858931} and \glspl*{hlc}~\cite{Matsuda2020, 7858291, 7918333, 8354004, 34567283647389, Jimenez-Naharro2017-ni, lyg2359, 23428376545267834}.
Due to the higher abstraction of \glspl*{ilc}, these protect against a certain Fault Model, whereas the \glspl*{hlc} are generally deployed to protect against a certain type of \gls*{fi}.
The last column determines, whether or not this countermeasure is theoretically able to protect from voltage \gls*{mfi} attacks, as described throughout our work.
A check (\checkmark) indicates, that the respective countermeasure is able to protect from multiple voltage fault injection, whereas a cross (×) indicates, that it is not.
Regarding the \glspl*{ilc}, these must be directed against Instruction Skipping in order to protect from our proposed \gls*{mfi} attack, whereas the \glspl*{hlc} must be deployed in order to detect \gls*{vfi}.

Moro et al.~\cite{2014} proposed a duplication based \gls*{ilc} replacement approach for most Thumb-2 instructions.
Moreover, Barry et al.~\cite{10.1145/2858930.2858931} proposed a follow-up \gls*{ilc}, as not every instruction could be automatically replaced by the modified \gls*{llvm} based compiler and hence, required manual analysis.
The authors base their countermeasures on the assumption that attacking successive instructions using \gls*{fi} is hard. We showed in \autoref{sec:successive_instructions} that, in principal, \coolname is able to attack successive instructions. Therefore, we denote this with a checkmark in parenthesis in \autoref{tab:countermeasures}.

Vosoughi et al.~\cite{lyg2359} propose a \gls*{hlc} which is able to mitigate the effect of \gls*{vfi} on the \gls*{ic}. Naturally, it can also protect against \gls*{mfi} using \gls*{vfi}, but it depends on specific on-chip voltage regulators to be present in the \gls*{ic}, which is not present at all times. Similarly, Singh et al.~\cite{23428376545267834} propose an application specific \gls*{hlc}, which has to be adapted to the specific application to be protected.

\section{Conclusion}\glsresetall
In this paper, we introduced a novel multiple voltage fault injection platform, coined \coolname, which is capable of injecting multiple, coordinated voltage faults into arbitrary target devices, in order to attack multiple fault targets during a single execution of the target's firmware.
We proposed and evaluated a novel, efficient parameter search algorithm for multiple voltage fault injection attacks.
By the hereby introduced attack vector, a novel threat model emerges, in which the adversary is capable of defeating most instruction level countermeasures, as they are mostly implemented to protect from single fault injection attacks.
We have shown, that by using our novel approach a TrustZone-M implementation can be attacked, in which there are multiple, inter-dependent fault targets to overcome, including a specific fault injection protection.
Finally, we have discussed possible countermeasures to thwart multiple fault injection attacks.

\bibliographystyle{plain}
\bibliography{bibliography}

\begin{thebibliography}{10}

\bibitem{saml11}
Atmel.
\newblock {\em SAM L10/L11 Family Datasheet}, 2019.

\bibitem{aumuller2002fault}
Christian Aum{\"u}ller, Peter Bier, Wieland Fischer, Peter Hofreiter, and J-P
  Seifert.
\newblock Fault attacks on rsa with crt: Concrete results and practical
  countermeasures.
\newblock In {\em International Workshop on Cryptographic Hardware and Embedded
  Systems}. Springer, 2002.

\bibitem{5412860}
Alessandro Barenghi, Guido Bertoni, Emanuele Parrinello, and Gerardo Pelosi.
\newblock Low voltage fault attacks on the rsa cryptosystem.
\newblock In {\em 2009 Workshop on Fault Diagnosis and Tolerance in
  Cryptography (FDTC)}, 2009.

\bibitem{10.1145/1873548.1873555}
Alessandro Barenghi, Luca Breveglieri, Israel Koren, Gerardo Pelosi, and
  Francesco Regazzoni.
\newblock Countermeasures against fault attacks on software implemented aes:
  Effectiveness and cost.
\newblock In {\em Proceedings of the 5th Workshop on Embedded Systems
  Security}, WESS '10. Association for Computing Machinery, 2010.

\bibitem{10.1145/2858930.2858931}
Thierno Barry, Damien Courouss\'{e}, and Bruno Robisson.
\newblock Compilation of a countermeasure against instruction-skip fault
  attacks.
\newblock In {\em Proceedings of the Third Workshop on Cryptography and
  Security in Computing Systems}, CS2 '16. Association for Computing Machinery,
  2016.

\bibitem{10.1007/11502760_24}
Eli Biham, Louis Granboulan, and Phong~Q. Nguyen.
\newblock Impossible fault analysis of rc4 and differential fault analysis of
  rc4.
\newblock In Henri Gilbert and Helena Handschuh, editors, {\em Fast Software
  Encryption}. Springer Berlin Heidelberg, 2005.

\bibitem{10.1007/BFb0052259}
Eli Biham and Adi Shamir.
\newblock Differential fault analysis of secret key cryptosystems.
\newblock In Burton~S. Kaliski, editor, {\em Advances in Cryptology --- CRYPTO
  '97}. Springer Berlin Heidelberg, 1997.

\bibitem{6976638}
Johannes Blömer, Ricardo Gomes~da Silva, Peter Günther, Juliane Krämer, and
  Jean-Pierre Seifert.
\newblock A practical second-order fault attack against a real-world pairing
  implementation.
\newblock In {\em 2014 Workshop on Fault Diagnosis and Tolerance in
  Cryptography}, 2014.

\bibitem{Bozzato_Focardi_Palmarini_2019}
Claudio Bozzato, Riccardo Focardi, and Francesco Palmarini.
\newblock Shaping the glitch: Optimizing voltage fault injection attacks.
\newblock {\em IACR Transactions on Cryptographic Hardware and Embedded
  Systems}, 2019(2), 2019.

\bibitem{7918333}
Jakub Breier, Shivam Bhasin, and Wei He.
\newblock An electromagnetic fault injection sensor using hogge phase-detector.
\newblock In {\em 2017 18th International Symposium on Quality Electronic
  Design (ISQED)}, 2017.

\bibitem{10.1145/3460120.3484779}
Robert Buhren, Hans-Niklas Jacob, Thilo Krachenfels, and Jean-Pierre Seifert.
\newblock One glitch to rule them all: Fault injection attacks against amd's
  secure encrypted virtualization.
\newblock In {\em Proceedings of the 2021 ACM SIGSAC Conference on Computer and
  Communications Security}, CCS '21. Association for Computing Machinery, 2021.

\bibitem{10.1007/978-3-319-08302-5_16}
Rafael~Boix Carpi, Stjepan Picek, Lejla Batina, Federico Menarini, Domagoj
  Jakobovic, and Marin Golub.
\newblock Glitch it if you can: Parameter search strategies for successful
  fault injection.
\newblock In Aur{\'e}lien Francillon and Pankaj Rohatgi, editors, {\em Smart
  Card Research and Advanced Applications}. Springer International Publishing,
  2014.

\bibitem{10.1007/978-3-540-77048-0_22}
Hua Chen, Wenling Wu, and Dengguo Feng.
\newblock Differential fault analysis on clefia.
\newblock In Sihan Qing, Hideki Imai, and Guilin Wang, editors, {\em
  Information and Communications Security}. Springer Berlin Heidelberg, 2007.

\bibitem{263816}
Zitai Chen, Georgios Vasilakis, Kit Murdock, Edward Dean, David Oswald, and
  Flavio~D. Garcia.
\newblock Voltpillager: Hardware-based fault injection attacks against intel
  {SGX} enclaves using the {SVID} voltage scaling interface.
\newblock In {\em {USENIX} Security}. {USENIX} Association, 2021.

\bibitem{10.1007/978-3-030-97348-3_9}
Brice Colombier, Paul Grandamme, Julien Vernay, {\'E}milie Chanavat, Lilian
  Bossuet, Lucie de~Laulani{\'e}, and Bruno Chassagne.
\newblock Multi-spot laser fault injection setup: New possibilities for fault
  injection attacks.
\newblock In Vincent Grosso and Thomas P{\"o}ppelmann, editors, {\em Smart Card
  Research and Advanced Applications}. Springer International Publishing, 2022.

\bibitem{206182}
Ang Cui and Rick Housley.
\newblock {BADFET}: Defeating modern secure boot using second-order pulsed
  electromagnetic fault injection.
\newblock In {\em 11th {USENIX} Workshop on Offensive Technologies ({WOOT}
  17)}. {USENIX} Association, 2017.

\bibitem{8354004}
Chinmay Deshpande, Bilgiday Yuce, Leyla Nazhandali, and Patrick Schaumont.
\newblock Employing dual-complementary flip-flops to detect emfi attacks.
\newblock In {\em 2017 Asian Hardware Oriented Security and Trust Symposium
  (AsianHOST)}, 2017.

\bibitem{10.1007/978-3-540-45203-4_23}
Pierre Dusart, Gilles Letourneux, and Olivier Vivolo.
\newblock Differential fault analysis on a.e.s.
\newblock In Jianying Zhou, Moti Yung, and Yongfei Han, editors, {\em Applied
  Cryptography and Network Security}. Springer Berlin Heidelberg, 2003.

\bibitem{7858291}
Wei He, Jakub Breier, Shivam Bhasin, Noriyuki Miura, and Makoto Nagata.
\newblock An fpga-compatible pll-based sensor against fault injection attack.
\newblock In {\em 2017 22nd Asia and South Pacific Design Automation Conference
  (ASP-DAC)}, 2017.

\bibitem{10.1007/978-3-540-71039-4_10}
Michal Hojs{\'i}k and Bohuslav Rudolf.
\newblock Differential fault analysis of trivium.
\newblock In Kaisa Nyberg, editor, {\em Fast Software Encryption}. Springer
  Berlin Heidelberg, 2008.

\bibitem{10.1145/3152701.3152709}
Yeongjin Jang, Jaehyuk Lee, Sangho Lee, and Taesoo Kim.
\newblock Sgx-bomb: Locking down the processor via rowhammer attack.
\newblock In {\em Proceedings of the 2nd Workshop on System Software for
  Trusted Execution}, SysTEX'17. Association for Computing Machinery, 2017.

\bibitem{Jimenez-Naharro2017-ni}
Ra{\'u}l Jim{\'e}nez-Naharro, Fernando G{\'o}mez-Bravo, Jonathan
  Medina-Garc{\'\i}a, Manuel S{\'a}nchez-Raya, and Juan~Antonio
  G{\'o}mez-Gal{\'a}n.
\newblock A smart sensor for defending against clock glitching attacks on the
  {I2C} protocol in robotic applications.
\newblock {\em Sensors (Basel)}, 17(4), 2017.

\bibitem{251590}
Zijo Kenjar, Tommaso Frassetto, David Gens, Michael Franz, and Ahmad-Reza
  Sadeghi.
\newblock {V0LTpwn}: Attacking x86 processor integrity from software.
\newblock In {\em 29th USENIX Security Symposium (USENIX Security 20)}. USENIX
  Association, 2020.

\bibitem{kudera2018design}
Christian Kudera, Markus Kammerstetter, Markus M{\"u}llner, Daniel Burian, and
  Wolfgang Kastner.
\newblock Design and implementation of a negative voltage fault injection
  attack prototype.
\newblock In {\em 2018 IEEE International Workshop on Physical Attacks and
  Inspection of Electronics (PAINE)}, 2018.

\bibitem{203696}
Jaehyuk Lee, Jinsoo Jang, Yeongjin Jang, Nohyun Kwak, Yeseul Choi, Changho
  Choi, Taesoo Kim, Marcus Peinado, and Brent~ByungHoon Kang.
\newblock Hacking in darkness: Return-oriented programming against secure
  enclaves.
\newblock In {\em 26th USENIX Security Symposium (USENIX Security 17)}. USENIX
  Association, 2017.

\bibitem{LI20083727}
Wei Li, Dawu Gu, and Juanru Li.
\newblock Differential fault analysis on the aria algorithm.
\newblock {\em Information Sciences}, 178(19), 2008.

\bibitem{7774477}
Pei Luo, Yunsi Fei, Liwei Zhang, and A.~Adam Ding.
\newblock Differential fault analysis of sha3-224 and sha3-256.
\newblock In {\em 2016 Workshop on Fault Diagnosis and Tolerance in
  Cryptography (FDTC)}, 2016.

\bibitem{7004182}
P.~Maistri, R.~Leveugle, L.~Bossuet, A.~Aubert, V.~Fischer, B.~Robisson,
  N.~Moro, P.~Maurine, J.-M. Dutertre, and M.~Lisart.
\newblock Electromagnetic analysis and fault injection onto secure circuits.
\newblock In {\em 2014 22nd International Conference on Very Large Scale
  Integration (VLSI-SoC)}, 2014.

\bibitem{maldini2018genetic}
Antun Maldini, Niels Samwel, Stjepan Picek, and Lejla Batina.
\newblock Genetic algorithm-based electromagnetic fault injection.
\newblock In {\em 2018 Workshop on Fault Diagnosis and Tolerance in
  Cryptography (FDTC)}. IEEE, 2018.

\bibitem{Matsuda2020}
Kohei Matsuda, Sho Tada, Makoto Nagata, Yuichi Komano, Yang Li, Takeshi
  Sugawara, Mitsugu Iwamoto, Kazuo Ohta, Kazuo Sakiyama, and Noriyuki Miura.
\newblock An ic-level countermeasure against laser fault injection attack by
  information leakage sensing based on laser-induced opto-electric bulk current
  density.
\newblock {\em Japanese Journal of Applied Physics}, 59(SG), 2020.

\bibitem{34567283647389}
Noriyuki Miura, Zakaria Najm, Wei He, Shivam Bhasin, Xuan~Thuy Ngo, Makoto
  Nagata, and Jean-Luc Danger.
\newblock Pll to the rescue: a novel em fault countermeasure.
\newblock In {\em 2016 53nd ACM/EDAC/IEEE Design Automation Conference (DAC)}.
  IEEE, 2016.

\bibitem{2014}
N.~Moro, K.~Heydemann, E.~Encrenaz, and B.~Robisson.
\newblock Formal verification of a software countermeasure against instruction
  skip attacks.
\newblock {\em Journal of Cryptographic Engineering}, 4(3), 2014.

\bibitem{6623558}
Nicolas Moro, Amine Dehbaoui, Karine Heydemann, Bruno Robisson, and Emmanuelle
  Encrenaz.
\newblock Electromagnetic fault injection: Towards a fault model on a 32-bit
  microcontroller.
\newblock In {\em 2013 Workshop on Fault Diagnosis and Tolerance in
  Cryptography}, 2013.

\bibitem{8708249}
Onur Mutlu and Jeremie~S. Kim.
\newblock Rowhammer: A retrospective.
\newblock {\em IEEE Transactions on Computer-Aided Design of Integrated
  Circuits and Systems}, 39(8), 2020.

\bibitem{Nashimoto2017}
Shoei Nashimoto, Naofumi Homma, Yu-ichi Hayashi, Junko Takahashi, Hitoshi Fuji,
  and Takafumi Aoki.
\newblock Buffer overflow attack with multiple fault injection and a proven
  countermeasure.
\newblock {\em Journal of Cryptographic Engineering}, 7(1), 2017.

\bibitem{Nashimoto_Suzuki_Ueno_Homma_2021}
Shoei Nashimoto, Daisuke Suzuki, Rei Ueno, and Naofumi Homma.
\newblock Bypassing isolated execution on risc-v using side-channel-assisted
  fault-injection and its countermeasure.
\newblock {\em IACR Transactions on Cryptographic Hardware and Embedded
  Systems}, 2022(1), 2021.

\bibitem{ning2019understanding}
Zhenyu Ning and Fengwei Zhang.
\newblock Understanding the security of arm debugging features.
\newblock In {\em 2019 IEEE Symposium on Security and Privacy (SP)}, pages
  602--619. IEEE, 2019.

\bibitem{lpc55s69}
NXP.
\newblock {\em LPC55S6x/LPC55S2x/LPC552x User manual}, 2021.
\newblock Rev. 2.3.

\bibitem{rt6xx}
NXP.
\newblock {\em UM11147}, 2021.
\newblock Rev. 1.4.

\bibitem{o2016fault}
Colin O'Flynn.
\newblock Fault injection using crowbars on embedded systems.
\newblock {\em IACR Cryptol. ePrint Arch.}, 2016, 2016.

\bibitem{238594}
Colin O{\textquoteright}Flynn.
\newblock Min()imum failure: {EMFI} attacks against {USB} stacks.
\newblock In {\em {USENIX} Workshop on Offensive Technologies ({WOOT})}.
  {USENIX} Association, 2019.

\bibitem{cryptoeprint:2020:937}
Colin O'Flynn.
\newblock Bam bam!! on reliability of emfi for in-situ automotive ecu attacks.
\newblock Cryptology ePrint Archive, Report 2020/937, 2020.

\bibitem{cryptoeprint:2021:1217}
Colin O'Flynn.
\newblock Emfi for safety-critical testing of automotive systems.
\newblock Cryptology ePrint Archive, Report 2021/1217, 2021.

\bibitem{o2014chipwhisperer}
Colin O’Flynn and Zhizhang~David Chen.
\newblock Chipwhisperer: An open-source platform for hardware embedded security
  research.
\newblock In {\em International Workshop on Constructive Side-Channel Analysis
  and Secure Design}, pages 243--260. Springer, 2014.

\bibitem{10.1145/3427384.3427394}
Pengfei Qui, Dongsheng Wang, Yongqiang Lyu, and Gang Qu.
\newblock Voltjockey: Abusing the processor voltage to break arm trustzone.
\newblock {\em GetMobile: Mobile Comp. and Comm.}, 24(2), 2020.

\bibitem{123698251376}
Thomas Roth.
\newblock {\em TrustZone-M(eh): Breaking ARMv8-M's security}.
\newblock CCC, 2019.

\bibitem{10.1145/3319535.3354197}
Keegan Ryan.
\newblock Hardware-backed heist: Extracting ecdsa keys from qualcomm's
  trustzone.
\newblock In {\em Proceedings of the 2019 ACM SIGSAC Conference on Computer and
  Communications Security}, CCS '19. Association for Computing Machinery, 2019.

\bibitem{Sakamoto2021}
Junichi Sakamoto, Shungo Hayashi, Daisuke Fujimoto, and Tsutomu Matsumoto.
\newblock Constructing software countermeasures against instruction
  manipulation attacks: an approach based on vulnerability evaluation using
  fault simulator.
\newblock {\em Cluster Computing}, 2021.

\bibitem{10.1145/3338508.3359577}
Bodo Selmke, Florian Hauschild, and Johannes Obermaier.
\newblock Peak clock: Fault injection into pll-based systems via clock
  manipulation.
\newblock In {\em Proceedings of the 3rd ACM Workshop on Attacks and Solutions
  in Hardware Security Workshop}, ASHES'19. Association for Computing
  Machinery, 2019.

\bibitem{23428376545267834}
Arvind Singh, Monodeep Kar, Nikhil Chawla, and Saibal Mukhopadhyay.
\newblock Mitigating power supply glitch based fault attacks with fast
  all-digital clock modulation circuit.
\newblock In {\em 2019 Design, Automation \& Test in Europe Conference \&
  Exhibition (DATE)}, 2019.

\bibitem{soleimany2021practical}
Hadi Soleimany, Nasour Bagheri, Hosein Hadipour, Prasanna Ravi, Shivam Bhasin,
  and Sara Mansouri.
\newblock Practical multiple persistent faults analysis.
\newblock {\em Cryptology ePrint Archive}, 2021.

\bibitem{stm32l5}
STM.
\newblock {\em Arm TrustZone features for STM32L5 and STM32U5 Series}, 2021.
\newblock Rev. 5.

\bibitem{203864}
Adrian Tang, Simha Sethumadhavan, and Salvatore Stolfo.
\newblock {CLKSCREW}: Exposing the perils of {Security-Oblivious} energy
  management.
\newblock In {\em 26th USENIX Security Symposium (USENIX Security 17)}, pages
  1057--1074, Vancouver, BC, August 2017. USENIX Association.

\bibitem{8167704}
Niek Timmers and Cristofaro Mune.
\newblock Escalating privileges in linux using voltage fault injection.
\newblock In {\em 2017 Workshop on Fault Diagnosis and Tolerance in
  Cryptography (FDTC)}, 2017.

\bibitem{7774479}
Niek Timmers, Albert Spruyt, and Marc Witteman.
\newblock Controlling pc on arm using fault injection.
\newblock In {\em 2016 Workshop on Fault Diagnosis and Tolerance in
  Cryptography (FDTC)}, 2016.

\bibitem{5577278}
Elena Trichina and Roman Korkikyan.
\newblock Multi fault laser attacks on protected crt-rsa.
\newblock In {\em 2010 Workshop on Fault Diagnosis and Tolerance in
  Cryptography}, 2010.

\bibitem{10.1007/978-3-642-21040-2_15}
Michael Tunstall, Debdeep Mukhopadhyay, and Subidh Ali.
\newblock Differential fault analysis of the advanced encryption standard using
  a single fault.
\newblock In Claudio~A. Ardagna and Jianying Zhou, editors, {\em Information
  Security Theory and Practice. Security and Privacy of Mobile Devices in
  Wireless Communication}. Springer Berlin Heidelberg, 2011.

\bibitem{van2021fill}
Jan Van~den Herrewegen, David Oswald, Flavio~D Garcia, and Qais Temeiza.
\newblock Fill your boots: Enhanced embedded bootloader exploits via fault
  injection and binary analysis.
\newblock {\em IACR Transactions on Cryptographic Hardware and Embedded
  Systems}, pages 56--81, 2021.

\bibitem{6076471}
Jasper~G.J. van Woudenberg, Marc~F. Witteman, and Federico Menarini.
\newblock Practical optical fault injection on secure microcontrollers.
\newblock In {\em 2011 Workshop on Fault Diagnosis and Tolerance in
  Cryptography}, 2011.

\bibitem{10.1007/978-3-030-15462-2_12}
Sebastian Vasile, David Oswald, and Tom Chothia.
\newblock Breaking all the things---a systematic survey of firmware extraction
  techniques for iot devices.
\newblock In Beg{\"u}l Bilgin and Jean-Bernard Fischer, editors, {\em Smart
  Card Research and Advanced Applications}. Springer International Publishing,
  2019.

\bibitem{lyg2359}
Ali Vosoughi and Sel\c{c}uk K\"{o}se.
\newblock Leveraging on-chip voltage regulators against fault injection
  attacks.
\newblock In {\em Proceedings of the 2019 on Great Lakes Symposium on VLSI},
  GLSVLSI '19. Association for Computing Machinery, 2019.

\bibitem{9237302}
Vincent Werner, Laurent Maingault, and Marie-Laure Potet.
\newblock An end-to-end approach for multi-fault attack vulnerability
  assessment.
\newblock In {\em 2020 Workshop on Fault Detection and Tolerance in
  Cryptography (FDTC)}, 2020.

\bibitem{8167705}
Nils Wiersma and Ramiro Pareja.
\newblock Safety != security: On the resilience of asil-d certified
  microcontrollers against fault injection attacks.
\newblock In {\em 2017 Workshop on Fault Diagnosis and Tolerance in
  Cryptography (FDTC)}, 2017.

\bibitem{wu2020fast}
Lichao Wu, Gerard Ribera, Noemie Beringuier-Boher, and Stjepan Picek.
\newblock A fast characterization method for semi-invasive fault injection
  attacks.
\newblock In {\em Cryptographers’ Track at the RSA Conference}. Springer,
  2020.

\bibitem{yuce2018fault}
Bilgiday Yuce, Patrick Schaumont, and Marc Witteman.
\newblock Fault attacks on secure embedded software: Threats, design, and
  evaluation.
\newblock {\em Journal of Hardware and Systems Security}, 2(2), 2018.

\end{thebibliography}
\section{Appendix}\glsresetall

\subsection{Simulation 1}
\begin{figure}[h!]
    \begin{lstlisting}[
        language=C, 
        caption={Simulation of Duplicate Registers: Non-distinguishable Fault Targets},
        label={lst:parameter_non_cooperative}
        ]
enum res_t = {FAILURE, SUCCESS}
uint32_t some_register    = 0x00000000;
uint32_t some_register_DP = 0x00000000;
extern uint32_t reg_value;

res_t experiment_DR(void) {
    DELAY_1
    some_register    = reg_value;
    DELAY_2
    some_register_DP = reg_value;
    DELAY_3
    
    // In our setup, we made sure not to 
    // glitch the if-statements
    if (!some_register && !some_register_DP) {
        return SUCCESS;
    }
    return FAILURE;
}
\end{lstlisting}
\end{figure}

\subsection{Simulation 2}
\begin{figure}[h!]
\begin{lstlisting}[
    language=C, 
    caption={Simulation of Duplicate Registers: Distinguishable Fault Targets},
    label={lst:cooperative_fault_target}
    ]
enum res_t = {FAILURE, FIRST, SECOND, SUCCESS}
uint32_t some_register    = 0x00000000;
uint32_t some_register_DP = 0x00000000;
extern uint32_t reg_value;

res_t experiment_DR(void) {
    set_trigger()  ; // synchronization
    reset_trigger(); // purposes
    
    DELAY_1
    some_register = reg_value;
    DELAY_2
    some_register_DP = reg_value;
    DELAY_3
    
    // In our setup, we made sure not to 
    // glitch the if-statements
    if (!some_register && !some_register_DP)
        return SUCCESS;
    if (!some_register && some_register_DP)
        return FIRST;
    if (some_register && !some_register_DP)
        return SECOND;
    else
        return FAILURE;
}
\end{lstlisting}
\end{figure}

\subsection{Attacking Successive Instructions}
\begin{table*}[h!]
    \centering
    \begin{tabular}{|c|l|l|l|l|l|}
    \hline
    \#Faults & \multicolumn{4}{l|}{\hfil Result Distribution} & Invalid\\\hline
    & None (a=0x12) & Only LSLS (a=0x9) &  Only LSRS (a=0x38) & Both (a=0x13) & \\\hline
    Single wide fault        & 0.31 & 0.09 & 0.17 & 0.24 & 0.19 \\\hline
    Two narrow faults         & 0.17 & 0.19 & 0.21 & 0.15 & 0.28 \\\hline
    \end{tabular}
    \centering
    \caption{Distribution of results when simulating the privilege escalation for either a single, wider fault or two, narrow ones to be injected in order to attack directly successive instructions.}
    \label{table:results_succeeding_fault_targets}
\end{table*}

\subsection{Evading Brown-out-Detection}
\begin{figure}[h!]
    \centering
    \includegraphics[width=0.45\textwidth]{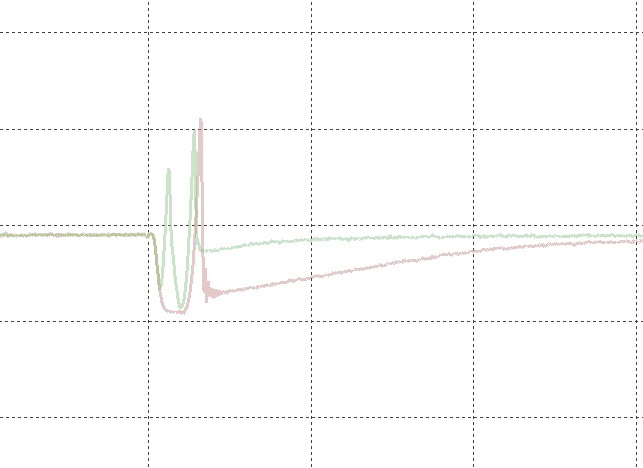}
    \caption{Oscilloscope recording of the original, wider glitch in red and its corresponding double-glitch counterpart in green. $2 \mu s$ per division}
    \label{fig:splitglitch}
\end{figure}

\subsection{Transferability}
\begin{table*}[]
\centering
\begin{tabular}{lccccrrr}
\hline
Chip     & FT1 & FT2                    & FT3 & FT4  & \begin{tabular}[c]{@{}r@{}}Exhaustive\\ Search\end{tabular} & \begin{tabular}[c]{@{}r@{}}Sweeping\\ Search\end{tabular} & Successrate \\ \hline
LPC55SXX & SAU & Secure AHB CTRL        & DR  & BXNS & \textgreater{}48h                                           & 8.15h                                                     & 0.0000003   \\
RT6XX    & SAU & Secure AHB CTRL        & DR  & BXNS & \textgreater{}48h                                           & 8.34h                                                     & 0.0000002   \\
SAML11   & SAU & Bootloader*             & -   & BXNS & \textgreater{}48h                                           & 4h                                                        & 0.0002566   \\
STM32L5  & SAU & Gobal TZ Configuration & -   & BXNS & \textgreater{}48h                                            & 47h                                                        & 0.0034000        \\ \hline
\end{tabular}

\caption*{*TZ Configuration stored in Non-Volatile-Memory (NVM), which is loaded by bootloader}
\caption{Evaluation of Multiple Voltage Fault Injection (MVFI) attacks on the TrustZone-M (TZM) implementation of different chips and the Fault Targets to be hit.}
\label{tab:chips}
\end{table*}

\subsection{Countermeasures}
\begin{table*}[h!]
    \centering
    \begin{tabular}{|c|c|c|c|c|c|}
        \hline
        Reference    & ILC   & HLC   & Fault Model / FI Method &  Protects against VMFI \\\hline
        Duplicate Registers~\cite{lpc55s69, rt6xx} & × & & Instruction Skipping & × \\\hline
        Barenghi et al.~\cite{10.1145/1873548.1873555} & × & & Instruction Skipping & × \\\hline
        Sakamoto et al.~\cite{Sakamoto2021} & × & & Instruction Manipulation & × \\\hline
        Moro et al.~\cite{2014} & × & & Instruction Skipping & (\checkmark) \\\hline
        Barry et al.~\cite{10.1145/2858930.2858931} & × & & Instruction Skipping & (\checkmark) \\\hline
        Matsuda et al.~\cite{Matsuda2020} & & × & \gls*{lfi} & × \\\hline
        Wei et al.~\cite{7858291} & & × & \gls*{lfi} & × \\\hline
        Breier et al.~\cite{7918333} & & × & \gls*{emfi} & × \\\hline
        Deshpande et al.~\cite{8354004} & & × & \gls*{emfi} & × \\\hline
        Miura et al.~\cite{34567283647389} & & × & \gls*{emfi} & × \\\hline
        Jimenez-Naharro et al.~\cite{Jimenez-Naharro2017-ni} & & × & \gls*{cfi} & × \\\hline
        Vosoughi et al.~\cite{lyg2359} & & × & \gls*{vfi} & \checkmark \\\hline
        Singh et al.~\cite{23428376545267834} & & × & \gls*{vfi} & \checkmark* \\\hline
    \end{tabular}
    \caption*{*Application specific}
    \caption{List of general \glspl*{ilc} and \glspl*{hlc} and their capability to  protecting against voltage \gls*{mfi} attacks}
    \label{tab:countermeasures}
\end{table*}

\end{document}